\documentclass[showpacs,twocolumn,preprintnumbers,amsmath,amssymb]{revtex4-1}

\newcommand{\beq}{\begin{equation}}
\newcommand{\eeq}{\end{equation}}

\newcommand{\id} {i\kern.06em\hbox{\raise.25ex\hbox{$/$}\kern-.60em$\partial$}}

\newcommand{\DD}{{\bf D}}
\newcommand{\LDD}{\overleftarrow{\DD}}

\newcommand{\ns}{/\kern-.52em n}
\newcommand{\ks}{/\kern-.52em k}

\newcommand{\bs}{/\kern-.52em b}
\newcommand{\qs}{/\kern-.52em s}

\newcommand{\p}{\partial}

\newcommand{\dd}
{\kern.06em\hbox{\raise.25ex\hbox{$/$}\kern-.60em$\partial$}}

\newcommand{\bsigma}{\mbox{\boldmath$\sigma$}}

\newcommand{\lj}{\langle}
\newcommand{\rj}{\rangle}
\newcommand{\vep}{\varepsilon}
\newcommand{\bx}{\boldsymbol{x}}
\newcommand{\by}{\boldsymbol{y}}
\newcommand{\bk}{\boldsymbol{k}}

\newcommand{\bp}{\boldsymbol{p}}

\newcommand{\bpsi}{\overline{\psi}}
\newcommand{\bP}{\boldsymbol{P}}

\newcommand{\MyI}{\mathbb{I}}

\newcommand{\bO}{{\bf O}}

\usepackage{graphicx}
\usepackage{dcolumn}
\usepackage{bm}
\usepackage{mathrsfs}

\usepackage{multirow}
\usepackage{empheq}
\usepackage[theorems,skins]{tcolorbox}

\DeclareMathAlphabet{\mathpzc}{OT1}{pzc}{m}{it}

\input amssym.def
\input amssym

\input epsf

\begin{document}
\title{A  Generally Covariant Theory of  Quantized Dirac Field  in de Sitter Spacetime}
\author{Sze-Shiang Feng$^\dag$, Mogus Mochena}
\affiliation
{Physics Department, Florida A \& M  University, Tallahassee, FL 32307\\
 $^\dag$email:sshfeng2014@gmail.com}
\date{\today}

\baselineskip 0.17in
\begin{abstract}
As a sequel to our previous work\cite{Feng2020}, we propose in this paper a quantization scheme for Dirac field in de Sitter spacetime.  Our scheme is covariant under both general transformations and Lorentz transformations. We first present a Hamiltonian structure, then quantize the field following the standard approach of constrained systems.  For the free field, the time-dependent quantized Hamiltonian is diagonalized by Bogliubov transformation and the eigen-states at each instant are interpreted as the observed particle states at that instant. The measurable energy-momentum of observed particle/antiparticles are the same as obtained for Klein-Gordon field. Moreover, the energy-momentum also satisfies geodesic equation, a feature justifying its measurability. As in \cite{Feng2020}, though the mathematics is carried out in terms of conformal coordinates for the sake of convenience, the whole theory can be transformed into any other coordinates based on general covariance.  It is concluded that particle/antiparticle states, such as vacuum states in particular are time-dependent and vacuum states at one time evolves into non-vacuum states at later times.  Formalism of perturbational calculation is provided with an extended Dirac picture.
\end{abstract}

\maketitle

\section{Introduction}

As is well-known, one of the biggest challenges in theoretical physics is quantizing gravity, to which huge efforts have been paid since decades ago. Yet, the light of the other end of the tunnel is still not close, if not out of sight\cite{Smolin2002}. As a less ambitious middle step, quantizing all fields except gravity in curved spacetime has been attempted at  for decades\cite{Birrell}-\cite{Bar}. Nevertheless, most endeavors prior to\cite{Feng2020} ended up being not quite satisfactory, even for a particular curved spacetime, due to various reasons as discussed in\cite{Feng2020}. The importance of quantum field theory (QFT) in de Sitter spacetime has long been realized. Recent loop quantum gravity also suggests that QFT in de Sitter spacetime is a good approximation to the physics of gravity-matter system described by the product of gravitational Kodama Chern-Simons wave functional and the matter wave functional \cite{Smolin2002}, though the normalizability might impose a difficulty\cite{Witten2003}.  In our previous work\cite{Feng2020}, a real Klein-Gordon field is quantized with the help of vierbein and the  approach is generally covariant. In parallel with QFT  in Minkowski spacetime, Hamiltonian, particles states and perturbation theory were consistently defined. In particular, the measurable energy-momentum, which can be used to discuss cosmological redshift if the universe can be approximated by de Sitter spacetime, was also obtained. It is understood that the first quantization fields in curved spacetime can be implemented in a broad range of curved spacetimes whereas second quantization heavily depends the specific spacetime under consideration. The reason is that second quantization requires expansion in terms of field modes, which heavily depends on some convenient choice of coordinate system of the particular spacetime. Once the convenient set of basis solutions is available,  second quantization can be implemented and transformed to any other coordinate system. Hence the whole scheme is made generally covariant.  Not all sets of basis solutions are created equal since it is required that the basis can be interpreted as the states of a free particle. In Minkowski spacetime, that is the set of plane waves. 
Similar basis are also viable in de Sitter spacetime.

The key point in our approach is that the canonical conjugate of a generic field $\varphi$ in cured spacetime is the partial derivative of Lagrangian density with respect to the  time derivative projected onto the vierbein.   There are two benefits of this definition. First, it renders the definition generally covariant. Second, the "time" derivative is with respect to {\it measurable time} instead of  {\it coordinate time}. Suppose $\mathscr{D}_\mu$ is the covariant derivative (ie.g. generally for a scalar/vector or Lorentzian for a spinor) of $\varphi$, then the "time" derivative is $e^\mu_0 \mathscr{D}_\mu\varphi$ where $e^\mu_a$ is the vierbein field. 

Quantization of Dirac field in de Sitter spacetime was discussed in literatures\cite{Nachtmann1967}-\cite{Cotaescu2019a}. Most discussion on the same topic so far are not generally covariant, not consistent with the requirement of general covariance and hence not fulfilling the original purpose.  As is well-understood, when a field interacts with time-dependent  external field, the whole system will be non-conservative and the Hamiltonian, which usually governs the evolution of the system, is supposed to be in general time-dependent. The Hamiltonian  and particle-antiparticle states in \cite{Nachtmann1967} are constant.  In \cite{Candelas1975}, there is no appropriate quantization procedure and  there is no creation of particles. Yet, a free field in de Sitter spacetime is actually an assembly of time-dependent oscillators hence all the concepts therein such as Hamiltonian, particle/antiparticle states should be time-dependent as in \cite{Chernikov1967}-\cite{Struckmeier2001}.   As is well understood, external lines in Feynman diagrams represent observable free particles. Without this interpretation of external lines, the physical meaning of all calculations will be at least vague since the calculated amplitude does not tell about scattering.  The (loop) calculations in \cite{Miao2006}-\cite{Miao2012} are based on the assumption that the external lines of Feynman diagrams represents plane wave $e^{i\bk\cdot\bx}$ with a time-dependent factor. Actually, the free-particle state in de Sitter spacetime is a mixture of $e^{i\bk\cdot\bx}$ and 
 $e^{-i\bk\cdot\bx}$ , as has been understood \cite{Akhmedov2014}. Works \cite{Cotaescu2002}-\cite{Cotaescu2019a} mainly focus on the mathematical features/generation of solutions. The quantization procedure is not based on time-dependent Hamiltonian and hence the particle states are time-independent and the Green's functions do not describes the propagation of physical particle from one state at one time to another state at another time.
In this work, we present a fundamentally different approach to quantizing Dirac field in de Sitter spacetime.  As in\cite{Feng2020}, our approach will be not only generally covariant under coordinate transformations but also Lorentz covariant under local Lorentz transformations.

    The present paper is arranged as follows.  In section II, we present the canonical structure and quantization of free Dirac field in de Sitter spacetime following the standard approach for constrained systems. New fundamental are also introduced. In section III,, the system is quantized in Schr\"{o}dinger picture using the new fundamental fields. Section IV provides fundamental solutions of Dirac field in de Sitter spacetime, as a preparation of second quantization. Section V presents in detail the field 2nd quantization and the quantized Hamiltonian both in Heisenberg picture and Schr\"{o}dinger picture. The Hamiltonians are diagonalized and quasi-particle creation/annihilation operators are defined. Discussions of difference as well as thing in common in the two pictures are provided. Time dependent vacuum and particles states are defined. Particularly, the observed energy-momentum is obtained based on our previous work and the on-shell relation for free particles is obtained. Some simple matrix elements are calculated.  In section VI, we define a generation functional which can be used to calculate various matrix elements. In section VII, transition amplitude between two states at different times is formulated. Section VIII is devoted to formulation of perturbation theory for interacting field, with the help of Dirac picture.  Section IX is a discussion of local Lorentz covariance of our approach.  The last section X is conclusional discussion.    
    
\section{Canonical Structure and Quantization of Dirac Field in de Sitter Spacetime}
We use the same notational convention as in \cite{Feng2020}: $
ds^2=C(\zeta)[d\zeta^2-\sum_i (dx^i)^2]
$
with conformal factor $C(\zeta)=(\ell/\zeta)^2, x=(\zeta,\bx), \mu=\zeta, 1,2,3$ and $ i,j=1,2,3, g_{\mu\nu}=\eta_{ab}e^a_\mu e^\nu_b$. 
 The symmetric, real Lagrange of a free Dirac field is\cite{Itzykson1980}-\cite{Ryder1996}
 \beq
\mathscr{L}_\psi=\frac{i}{2}\Big[\bpsi \gamma^a e^\mu_a\DD_\mu\psi-\bpsi\LDD_\mu e^\mu_a\gamma^a\psi\Big]-m_\psi\bpsi\psi \label{symL}
\eeq
with $ \DD_\mu=\p_\mu+\frac{i}{4}\omega_{\mu mn}\Sigma^{mn}=\p_\mu-\frac{1}{4}\omega_{\mu mn}\gamma^m\gamma^n=\p_\mu-{\bf A}_\mu, {\bf A}_\mu=\frac{1}{4}\omega_{\mu mn}\gamma^m\gamma^n, \gamma^0{\bf A}_\mu^\dag\gamma^0=-{\bf A}_\mu,  \Sigma_{mn}=\frac{i}{2}[\gamma_m,\gamma_n], \bpsi \LDD_\mu:=\p_\mu\bpsi+\bpsi {\bf A}_\mu. $
 The Euler-Lagrangian equation $[e\mathscr{L}_\psi]_{\bpsi}=0$ is
\beq
\frac{i}{2}e\gamma^a e^\mu_a\DD_\mu\psi-\frac{i}{2}e{\bf A}_\mu e^\mu_a\gamma^a\psi-em_\psi\psi+\p_\mu (\frac{i}{2}ee^\mu_a\gamma^a\psi)=0
\eeq
Since
$
[{\bf A}_\mu,\gamma^a]=-\omega_\mu\,^{an}\gamma_n
$, then
$
{\bf A}_\mu e^\mu_a\gamma^a= e^\mu_a\gamma^a {\bf A}_\mu  +\omega_n\gamma^n
$
with $\omega_a=\nabla_\mu e^\mu_a$. 
Using $, \p_\mu(ee^\mu_a)=e\nabla_\mu e^\mu_a, \nabla_\mu e^\nu_b=\omega_{\mu b}\,^ce^\nu_c$, we have
\begin{eqnarray}
i\gamma^a e^\mu_a\DD_\mu\psi-m_\psi\psi&=&0\\
i\bpsi\LDD_\mu\gamma^a e^\mu_a+m_\psi\bpsi&=&0
\end{eqnarray}
The Dirac current is defined as
\beq
J^\mu_{\rm Dirac}(\psi,\varphi)=\bpsi\gamma^a e^\mu_a\varphi
\eeq
It can be easily verified that
$
\nabla_\mu J^\mu_{\rm Dirac}=0. 
$
The inner-product is defined
\begin{align}
\left(\psi\mid \varphi\right)_{\rm Dirac}:&=\int_\Sigma d\sigma_\mu \bpsi\gamma^a e^\mu_a\varphi=\int_\Sigma d\sigma \psi^\dag\varphi
\end{align}
So we have
$
\left(\psi\mid\varphi\right)^*_{\rm Dirac}=(\varphi\mid\psi)_{\rm Dirac}
$
which is general-Lorentz invariant.

Denoting $\hat{\DD}_a=e^\mu_a\DD_\mu,  \hat{\LDD}_a=e^\mu_a\LDD_\mu$,
 (here $e^\mu_a$ plays the role of parameter $\lambda(t)$ in \cite{Landau-Lifshitz1951}) .
The canonical conjugates are defined as \cite{Goldstein1980}
\beq
\pi_\psi:=\frac{\p\mathscr{L}_\psi}{\p\hat{\DD}_0\psi}=\frac{1}{2}i\bpsi\gamma^0,\,\,\,\,\,
 \pi_{\bpsi}:= \frac{\p\mathscr{L}_\psi}{\p\bpsi\hat{\LDD}_0}=-\frac{1}{2}i\gamma^0\psi
\eeq
Unlike the situation of Klein-Gordon field,  there does not exist the inverse 
$
  \hat{\DD}_0\psi= \hat{\DD}_0\psi(\psi, \hat{\DD}_i\psi,\pi_\psi,\bpsi,\bpsi\hat{\LDD}_i,\pi_{\bpsi};x).
$
Instead, the system is a constrained system with two primary constraints\cite{Gitman1990}
\begin{align}
\mathpzc{C}_1&=\pi_\psi-\frac{1}{2}i\bpsi\gamma^0\approx 0\\
\mathpzc{C}_2&=\pi_{\bpsi}+ \frac{1}{2}i\gamma^0\psi\approx 0
\end{align}
Hence the canonical Hamiltonians $\mathscr{H}$ and the total Hamiltonian $\mathscr{H}_T$ can be constructed via
\begin{align}
\mathscr{H}&=\bpsi\hat{\LDD}_0\pi_{\bpsi} +\pi_\psi \hat{\DD}_0\psi-\mathscr{L}_\psi\\
\mathscr{H}_T&=\mathscr{H}+\mathpzc{C}_1\lambda^1+\lambda^2 \mathpzc{C}_2
\end{align}
where $\lambda^i, i=1,2$ are odd variables. Classical Poisson brackets between the primary constraints $\mathpzc{C}_i$ and the total Hamiltonian $\mathscr{H}_T$ determine $\lambda^i$s completely. Consequently, there are no secondary constraints and the system has only primary second-class constrains.
The canonical Hamiltonian is expressed as
\begin{align}
&H[\bpsi,\psi;\pi_\psi,\pi_{\bpsi};\zeta]\nonumber\\
&=\int_\Sigma d\sigma\Big[
\frac{i}{2}(-\bpsi \gamma^{a'} \hat{\DD}_{a'}\psi+\bpsi\hat{\LDD}_{a'}\gamma^{a'}\psi)+m_\psi\bpsi\psi\Big]
\end{align}
and the total Hamiltonian is then
\begin{align}
&H_T[\bpsi,\psi;\pi_\psi,\pi_{\bpsi};\zeta]\nonumber\\
&=\int_\Sigma d\sigma\Big[
\frac{i}{2}(-\bpsi \gamma^{a'} \hat{\DD}_{a'}\psi+\bpsi\hat{\LDD}_{a'}\gamma^{a'}\psi)+m_\psi\bpsi\psi\nonumber\\
&
+\mathpzc{C}_1\lambda^1+\lambda^2 \mathpzc{C}_2\Big]
\end{align}
 Equal-time Poisson brackets are defined
\begin{align}
\left\{\psi_A(x),\pi_{\psi B}(y)\right\}^{\text{P.B.}}_{x^0=y^0}:&=\delta_{AB}\delta^3(Z^i)\\
\left\{\psi_A(x),\bpsi_B(y)\right\}^{\text{P.B.}}_{x^0=y^0}&=0\\
\left\{\bpsi_A(x),\pi_{\bpsi B}(y)\right\}^{\text{P.B.}}_{x^0=y^0}:&=\delta_{AB}\delta^3(Z^i)\\
\left\{\pi_{\psi A}(x),\pi_{\bpsi B}(y)\right\}^{\text{P.B.}}_{x^0=y^0}&=0
\end{align}
where $Z^i$ is defined as in\cite{Feng2020}. The Poisson bracket matrix of the two constraints is
\beq
\Theta=\left\{\mathpzc{C}_{iA}(x),\mathpzc{C}_{jB}(y)\right\}^{\text{P.B.}}_{x^0=y^0}=
\left(
\begin{array}{cccc}
0& i\gamma^{0}\\
-i\gamma^{0T}&0
\end{array}\right)\delta^3(Z^i)
\eeq
and the inverse is 
\beq
\Theta^{-1}=
\left(
\begin{array}{cccc}
0& i\gamma^{0T}\\
-i\gamma^{0}&0
\end{array}\right)\delta^3(Z^i)
\eeq
Hence the Dirac brackets are
\begin{eqnarray}
\left\{\psi_A,\bpsi_B(x)\right\}^{\text{D.B.}}_{x^0=y^0}=-i\gamma^0_{AB}\delta^3(Z^i)\label{DB}
\end{eqnarray}
We can choose a new set of canonical variables and constraints.
\begin{align}
\omega&=\frac{1}{2}\psi+i\gamma^0\pi_{\bpsi},
\pi_{\omega}=\pi_\psi+\frac{1}{2}i\bpsi\gamma^0\\
\Omega&=\frac{1}{2}\bpsi+i\pi_\psi\gamma^0,
\pi_{\Omega}=\pi_{\bpsi}+\frac{1}{2}i\gamma^0\psi
\end{align}
with inverse transformation
\begin{align}
\psi&=\omega-i\gamma^0\pi_\Omega,  \pi_\psi=\frac{1}{2}(\pi_\omega-i\Omega\gamma^0)\\
\bpsi&=-i(\pi_\omega\gamma^0+i\Omega),\pi_{\bpsi}=-\frac{i}{2}(\gamma^0\omega+i\pi_\Omega)
\end{align}
With these new variables, the Poisson brackets and Dirac brackets are equal. 
The physical Hamiltonian
\begin{align}
\mathscr{H}_{\text {phy}}&=-i(\pi_\omega\gamma^0+i\Omega)\left[\frac{i}{2}\Big[- \gamma^{a'} \hat{\DD}_{a'}+\hat{\LDD}_{a'}\gamma^{a'}\Big]+m_\psi\right]\nonumber\\
&\times (\omega-i\gamma^0\pi_\Omega)_{|\Omega=\pi_\Omega=0}\nonumber\\
&=-i\pi_\omega\gamma^0\left[\frac{i}{2}\Big[- \gamma^{a'} \hat{\DD}_{a'}+\hat{\LDD}_{a'}\gamma^{a'}\Big]+m_\psi\right]\omega
\end{align}
Still denoting $\omega$ as $\psi$, and $\pi_\omega$ as $i\bpsi\gamma^0$, $(\psi,\bpsi)$ are now a canonical pair. We arrive at the conventional expression of Hamiltonian 
\begin{align}
\mathscr{H}_{\text {phy}}[\psi,\bpsi;\zeta]
&=\bpsi\left[\frac{i}{2}\Big[- \gamma^{a'} \hat{\DD}_{a'}+\hat{\LDD}_{a'}\gamma^{a'}\Big]+m_\psi\right]\psi \label{phyH}
\end{align}
It might be worthwhile noting that this agrees with the Faddeev-Jackiw method for first-order Lagrangians\cite{Faddeev1988}.  The purpose of our approach is not just to quantize the system but also in a covariant way, which many literatures have failed to address/emphasize so far. 
The P.B. as eq.(\ref{DB}), with which we can reproduce the original equation: 
\begin{eqnarray}
&&\hat{\DD}_0\psi(x)=\left\{\psi,H_{\text{phy}}[\bpsi,\psi]\right\}^{\text{P.B.}}\nonumber\\
&=&\int_\Sigma d\sigma \left\{\psi(x),
\frac{i}{2}(-\bpsi \gamma^{a'} \hat{\DD}_{a'}\psi+\bpsi\hat{\LDD}_{a'}\gamma^{a'}\psi)+m_\psi\bpsi\psi\right\}^{\text{P.B.}}\nonumber\\
&=&-i\gamma^0(-i\gamma^{a'}\hat{\DD}_{a'}\psi+m_\psi\psi)
\end{eqnarray}
i.e.,
\beq
ie^\mu_0\gamma^0\DD_\mu\psi=-i\gamma^{a'}e^\mu_{a'}\DD_\mu\psi+m_\psi\psi \label{Dirac-eqR}
\eeq
Similarly
\beq
ie^\mu_0\bpsi(x)\LDD_\mu\gamma^0=-i\bpsi\LDD_\mu e^\mu_{a'}\gamma^{a'}-m_\psi\bpsi  \label{Dirac-eqL}
\eeq
  We are now facing a similar issue encountered in KG case, i.e., the rhs of eq.(\ref{DB}) depends on time $\zeta$, which will cause inconsistency when introducing Schr\"{o}dinger picture.  {\it We need to extract partial information of the fields $\psi,\bpsi$, information of which Schr\"{o}dinger picture can be well-defined. }
  For this, we introduce fields
  \beq
  \xi=e^{\Omega/2}\psi, \bar{\xi}=e^{\Omega/2}\bpsi
  \eeq
  These fields are not Dirac {\it spinors in curved spacetime} and they do not transform the way the original fields do under local Lorentz transformations. From these definitions, we have
    \begin{align}
  \hat{\DD}_0\xi
  &=\frac{1}{2}\gamma^0\gamma^ae^\mu_a(\DD_\mu \Omega)\xi-\gamma^0\gamma^{a'}\hat{\DD}_{a'}\xi-i\gamma^0 m_\psi \xi 
    \end{align}
 This can also be obtained as
 \beq
 \hat{\DD}_0\xi=\left\{\xi,H_{\text{phy}}[\psi,\bpsi;\zeta]\right\}^{\text{P.B.}}+\frac{1}{2}e^\zeta_0\p_\zeta \Omega \cdot \xi
  \eeq
 If we choose $\xi,\bar{\xi}$ as fundamental fields and define the Poisson bracket for any two functionals $F[\xi,\bar{\xi};\zeta]=\int_\Sigma  d\sigma\mathscr{F}, G[\xi,\bar{\xi};\zeta]=\int_\Sigma  d\sigma\mathscr{G}$ as
\begin{align}
&\left\{ F,G\right\}^{\text{P.B.}}\nonumber\\
&:=-i\gamma^0_{AB}\int_\Sigma d\sigma e^{\Omega} \left(\frac{\delta \mathscr{F}}{\delta\xi_A(x)}\frac{\delta \mathscr{G}}{\delta\bar{\xi}_B(x)}-\frac{\delta \mathscr{G}}{\delta\xi_A(x)}\frac{\delta \mathscr{F}}{\delta\bar{\xi}_B(x)}\right)
\end{align}
Since $\xi(x)=\int_{\Sigma} d\sigma' \delta^3(x-x')C^{-3/2}(\zeta)\xi(x'), \bar{\xi}(y)=\int_{\Sigma} d\sigma' \delta^3(y-y')C^{-3/2}(\zeta)\bar{\xi}(y')$, we have then
\beq
\left\{\xi(x),\bar{\xi}(y)\right\}^{\text{P.B.}}=-ie^{\Omega}\gamma^0\delta^3(Z^i)=-i\gamma^0\delta^3(\bx-\by)
\eeq
Thus with the Hamiltonian
\begin{align}
&\mathscr{H}_{\text {phy}}[\xi,\bar{\xi};\zeta]\nonumber\\
&=e^{-\Omega}\bar{\xi}\Big[\frac{i}{2}(- \gamma^{a'} \hat{\DD}_{a'}+\hat{\LDD}_{a'}\gamma^{a'})+m_\psi+i\gamma^0\frac{1}{2}C^{-1/2}\omega_0\Big]\xi
\label{xiH}
\end{align}  
we have
\begin{align}
\hat{\DD}_0\xi&=\{\xi, H_{\text{phy}}[\xi,\bar{\xi};\zeta]\}^{\text{P.B.}}
\end{align}
 In general, for any $O[\xi,\bar{\xi};\lambda(x)]$
\beq
O[\xi,\bar{\xi};\lambda(x)]=\int_\Sigma d\sigma  \mathscr{O}(\xi,\bar{\xi};\lambda(x))
\eeq
which is a function of $\zeta$ and a functional of $\xi,\bar{\xi}$
\begin{align}
 \hat{d}_0O[\xi,\bar{\xi};\lambda(x)](\zeta)& =\int_\Sigma e^\zeta_0(x)\p_\zeta(d\sigma)\mathscr{O}+\left\{O, H_{\text{phy}}\right\}^{\text{P.B.}}\nonumber\\
 &+\int_\Sigma d\sigma\frac{\delta  \mathscr{O}}{\delta\lambda(x)}\hat{\nabla}_0\lambda
  \end{align}
 where $\hat{d}_0$ is defined as $e^\zeta_0(\zeta)d/d\zeta$, bearing in mind that $\zeta=\text{const.}$ defines the surface $\Sigma$, i.e., the l.h.s. depends on the surface $\Sigma$. In particular,
\begin{align}
 \hat{d}_0H_{\text{phy}}[\xi,\bar{\xi};\cdots]=&\int_\Sigma e^\zeta_0(x)\p_\zeta(d\sigma)\mathscr{H}_{\text{phy}}\nonumber\\
 &-\int_\Sigma d\sigma \frac{\delta  \mathscr{H}_{\text{phy}}}{\delta e^\mu_a}\hat{\nabla}_0e^\mu_a
\end{align}
Quantization is realized by the correspondence
\beq
\{, \}^{\text{D.B.}}\rightarrow \frac{1}{i\hbar}\{,\}
\eeq
Hence we have anticommutator
\beq
\left\{\xi(x),\bar{\xi}(y)\right\}=\gamma^0\delta^3(\bx-\by)
\eeq
In the sequel, we write $H_{\text{phy}}$ as $H$ for brevity.

 \section{Schr\"{o}dinger Picture}
 Upon quantization, the classical canonical variables $\psi,\bpsi$ are replaced by operators $\hat{\psi},\hat{\bpsi}$  in a Hilbert space and Poisson brackets become anti-commutators.
   In standard QFT in Minkowski spacetime, the Hamiltonian is time-independent and three pictures can be utilized. Similarly, 
we can define Schr\"{o}dinger picture for $O=\xi,\bar{\xi}$
\begin{align}
O^{\text{S}}(\zeta)=&\hat{T}^{-1}e^{i\int^\zeta_\ell H[\xi,\bar{\xi};\eta]e^0_\zeta(\eta)d\eta}O (\zeta)\hat{T}e^{-i\int^\zeta_\ell H[\xi,\bar{\xi};\eta]e^0_\zeta(\eta)d\eta}
\end{align}
where $\hat{T}$ is the time-ordering operator define as  $\hat{T}\xi(\zeta_1)\xi(\zeta_2)=\xi(\zeta_1)\xi(\zeta_2)\theta(\zeta_2-\zeta_1)-\xi(\zeta_2)\xi(\zeta_1)\theta(\zeta_1-\zeta_2).$  It should be emphasized that here what is employed is $H[\xi,\bar{\xi};\eta]$ instead of $H[\psi,\bpsi;\eta]$. The two pictures agree at $\zeta=\ell$
\beq
\xi^{\text{S}}(\bx)=\xi(\ell,\bx)=\psi(\ell,\bx)
\eeq
 For Hamiltonian
\begin{align}
H^{\text{S}}(\zeta)=&\hat{T}^{-1}e^{i\int^\zeta_\ell H(\eta)e^0_\zeta(\eta)d\eta}H (\zeta)\hat{T}e^{-i\int^\zeta_\ell H(\eta)e^0_\zeta(\eta)d\eta}\nonumber\\
=&H[\xi^{\text{S}}(\bx),\bar{\xi}^{\text{S}}(\bx);e^\mu_a(x)](\zeta)
\end{align}
and $
H^{\text{S}}(\ell)=H(\ell)
$. 
 So time-independent $\xi^{\text{S}}(\bx), \bar{\xi}^{\text{S}}(\bx)$ play the roles of $x,p$  while $\xi(x),\bar{\xi}(x)$ play the role $x_\pm,p_\pm$ in \cite{Landovitz1979} :
\begin{align}
H(t)=&f(t)p^2/2m+g(t)\frac{1}{2}m\omega^2_0x^2,\nonumber\\
H_+(t)=&f(t)p_+^2(t)/2m+g(t)\frac{1}{2}m\omega^2_0x_+^2(t)\nonumber
\end{align}
As suggested in 
\begin{align}
dH(t)/dt=&\dot{f}(t)p^2/2m+\dot{g}(t)\frac{1}{2}m\omega^2_0x^2,\nonumber\\
dH_+(t)/dt=&\dot{f}(t)p_+^2(t)/2m+\dot{g}(t)\frac{1}{2}m\omega^2_0x_+^2(t)\nonumber
\end{align}
, though the initial condition $H(0)=H_+(0)$, we have $H(t)\not=H_+(t)$ since $p_+(t), x_+(t)$ depend on time $t$. The time-dependence of $H^{\text{S}}$ is
\begin{align}
i\hat{d}_0H^{\text{S}}(\zeta)
=ie^\zeta_0\frac{\p}{\p\zeta} H[\xi^{\text{S}}(\bx),\bar{\xi}^{\text{S}}(\bx);e^\mu_a(x)](\zeta)
\end{align}
Quantum anticommutator is 
\begin{align}
\{\hat{\xi}^{\text{S}}(\bx), \hat{\bar{\xi}}\,^{\text{S}}(\by)\}_{x^0=y^0}=\gamma^0\hbar\delta^3(\bx-\by)
\end{align}
Quantum mechanical Schr\"odinger  eq. for  wave-functional
\beq
i\hat{\DD}_0\Psi[\xi^{\text{S}}(\bx),\zeta]=H^{\text{S}}[\xi^{\text{S}}(\bx),\bar{\xi}\,^{\text{S}}(\bx);\zeta]\Psi[\xi^{\text{S}}(\bx),\zeta]
\eeq
where in $H^{\text{S}}$  (As in conventional quantum field theories, this is not unique!)
$
\xi_A^{\text{S}}(\bx)\mapsto \hbar \gamma^0_{AB}\delta/\delta\bar{\xi}^{\text{S}}_B(\bx)
$
So 
\begin{align}
&H^{\text{S}}(\zeta)\nonumber\\
&=\int _\Sigma d\sigma e^{-\Omega} \bar{\xi}^{\text{S}}(\bx)\left[-i\gamma^{a'} \hat{\DD}_{a'}+m_\psi+i\gamma^0\frac{1}{2}C^{-1/2}\omega_0\right]\nonumber\\&\times \hbar \gamma^0\delta/\delta\bar{\xi}^{\text{S}}(\bx) \label{SchHaPsi}
\end{align}
One can discuss wave functional based on this Hamiltonian. But this is left out of the main scope of the present paper.
\section{Fundamental Solutions for Free Field}
With choosing viebein as $e^a_\mu=\delta^a_\mu C^{1/2}(\zeta)$ and the Dirac representation of $\gamma$-matrices, the spin connections $\omega_{\mu ab}$ are
\begin{align}
\omega_{0ab}&=0, 
\omega_{1ab}=\left(\begin{array}{cccc}
0&\frac{1}{\zeta}&0&0\\
-\frac{1}{\zeta}&0&0&0\\
0&0&0&0\\
0&0&0&0
\end{array}\right)\\
\omega_{2ab}&=\left(\begin{array}{cccc}
0&0&\frac{1}{\zeta}&0\\
0&0&0&0\\
-\frac{1}{\zeta}&0&0&0\\
0&0&0&0\\
\end{array}\right),
\omega_{3ab}=\left(\begin{array}{cccc}
0&0&0&\frac{1}{\zeta}\\
0&0&0&0\\
0&0&0&0\\
-\frac{1}{\zeta}&0&0&0\\
\end{array}\right)
\end{align}
${\bf A}_0=0,{\bf A}_i=-\frac{1}{2\zeta}\gamma^0\gamma_a\delta^a_i$.  The Dirac's derivative is
$
\DD_0=\p_0,
\DD_i=\p_i+\frac{1}{2\zeta}\gamma^0\gamma_a\delta^a_i.
$
Hence, 
\begin{widetext}
\begin{align}
&i\gamma^ae^\mu_a\DD_\mu=\left(
\begin{array}{cccc}
 \frac{i (2\zeta \p_0 -3)}{2 \ell } & 0 & \frac{i \zeta\p_3  }{\ell } & \frac{\zeta(i \p_1+\p_2 ) }{\ell } \\
 0 & \frac{i (2 \zeta\p_0  -3)}{2 \ell } & \frac{i \zeta(\p_1+i \p_2)  }{\ell } & -\frac{i \zeta\p_3 }{\ell } \\
 -\frac{i \zeta\p_3 }{\ell } & -\frac{\zeta(i \p_1+\p_2)  }{\ell } & -\frac{i (2\zeta\p_0 -3)}{2 \ell } & 0 \\
 \frac{\zeta(\p_2-i \p_1)  }{\ell } & \frac{i \zeta\p_3  }{\ell } & 0 & -\frac{i (2\zeta \p_0 -3)}{2 \ell } \\
\end{array}
\right)
\end{align}
Let $\psi(x;\bk)=A_{\bk}\Theta_{\bk}(\zeta)e^{i\bk\cdot\bx}$, then Dirac equation reads
\begin{align}
\left(
\begin{array}{cccc}
 \frac{i (2\zeta \p_0 -3)}{2 \ell }-m_\psi & 0 & \frac{ \zeta k_3  }{\ell } & \frac{\zeta(k_1-ik_2 ) }{\ell } \\
 0 & \frac{i (2 \zeta\p_0  -3)}{2 \ell }-m_\psi & \frac{i \zeta(-ik_1+k_2)  }{\ell } & -\frac{ \zeta k_3 }{\ell } \\
 -\frac{\zeta k_3 }{\ell } & -\frac{\zeta(k_1-ik_2)  }{\ell } & -\frac{i (2\zeta\p_0 -3)}{2 \ell } -m_\psi& 0 \\
 \frac{\zeta(-ik_2-k_1)  }{\ell } & \frac{\zeta k_3  }{\ell } & 0 & -\frac{i (2\zeta \p_0 -3)}{2 \ell }-m_\psi \\
\end{array}
\right)\Theta_{\bk}(\zeta)=0
\end{align}
Denote $\Theta_{\bk}=(\varphi_{\bk},\chi_{\bk})^T$. Then
\begin{align}
[\frac{i (2\zeta \p_0 -3)}{2 \ell }-m_\psi ]\varphi_{\bk}+\left(\begin{array}{cc}
 \frac{ \zeta k_3  }{\ell } & \frac{\zeta(k_1-ik_2 ) }{\ell } \\
  \frac{i \zeta(-ik_1+k_2)  }{\ell } & -\frac{ \zeta k_3 }{\ell } \\
\end{array}\right)\chi_{\bk}(\zeta)&=0\\
\left(\begin{array}{cc}
 \frac{ \zeta k_3  }{\ell } & \frac{\zeta(k_1-ik_2 ) }{\ell } \\
  \frac{i \zeta(-ik_1+k_2)  }{\ell } & -\frac{ \zeta k_3 }{\ell } \\
\end{array}\right)\varphi_{\bk}(\zeta)+[\frac{i (2\zeta \p_0 -3)}{2 \ell }+m_\psi ]\chi_{\bk}&=0
\end{align}
Let $\varphi_{\bk}=\zeta^2f_{\bk}$
and  $\nu=1/2+i\ell m_\psi$, we have solution
\beq
f_{\bk}(\zeta)=H^{(1)}_{\nu}(k\zeta);  H^{(2)}_{\nu}(k\zeta);
\eeq
Given $\varphi_{\bk}(\zeta)$, we have
\beq
\chi_{\bk}(\zeta)
=\frac{i}{\zeta k}\hat{\bk}\cdot\bsigma[\zeta\p_0+\nu-2]\varphi_{\bk}
\eeq
Consider solution
\beq
\varphi_{\bk}(\zeta)=e^{\ell\pi m_\psi }k^2\zeta^2H^{(2)}_{\nu}(k\zeta)\left(\begin{array}{c}c_{1\bk}\\ c_{2\bk}\end{array}\right)
+\hat{\bk}\cdot\bsigma k^2\zeta^2H^{(1)}_{\nu}(k\zeta)\left(\begin{array}{c}d_{3\bk}\\ d_{4\bk}\end{array}\right)
\eeq
Then
\begin{align}
\chi_{\bk}(\zeta)
&=e^{\ell\pi m_\psi }i(k\zeta )^2H^{(2)}_{\nu-1}(k\zeta)\hat{\bk}\cdot\bsigma\left(\begin{array}{c}c_{1\bk}\\ c_{2\bk}\end{array}\right)
+i(k\zeta)^2 H^{(1)}_{\nu-1}(k\zeta)\left(\begin{array}{c}d_{3\bk}\\ d_{4\bk}\end{array}\right)
\end{align}
Thus
\beq
\psi_{\bk}(\zeta)=k^2\zeta^2K\left(\begin{array}{c}c_{\bk}\\ d_{\bk}\end{array}\right)
\eeq
where
\beq
K=\left(
\begin{array}{cc}
e^{\ell\pi m_\psi }H^{(2)}_{\nu}(k\zeta) &\hat{\bk}\cdot\bsigma H^{(1)}_{\nu}(k\zeta)\\
e^{\ell\pi m_\psi }iH^{(2)}_{\nu-1}(k\zeta)\hat{\bk}\cdot\bsigma &i H^{(1)}_{\nu-1}(k\zeta)
\end{array}
\right)
\eeq
 These basis solutions were provided in \cite{Nachtmann1967}.  Using recurrence relations of Bessel function, we have
$
 K^\dag K=\frac{4}{\pi k\zeta}.
$
Hence we have
$
\Theta^\dag_{\bk}(\zeta)\Theta_{\bk}(\zeta)=\frac{4}{\pi}k^3\zeta^3(c^\dag_{\bk}c_{\bk}+d^\dag_{\bk}d_{\bk}).
$
 Thus we have Dirac inner-product
 \beq
 (\psi(x;\bk)|\psi(x;\bk'))_{\text{Dirac}}=A_{\bk}^2\frac{4}{\pi}k^3\ell^3\delta_{\bk,\bk'}(c^\dag_{\bk}c_{\bk}+d^\dag_{\bk}d_{\bk})
  \eeq
, which suggests $A_{\bk}=(\frac{4}{\pi} k^3\ell^3)^{-1/2}$ with normalization $c^\dag_{\bk}c_{\bk}+d^\dag_{\bk}d_{\bk}=1$. We have for each $\bk$ 4 basis solutions $(c_{\bk},d_{\bk})^T: w^1=(1,0,0,0)^T, w^2=(0,1,0,0)^T, w^3=(0,0, 1,0)^T, w^4=(0,0,0,1)^T$.
\beq
\psi(x)=\sum_{\bk} A_{\bk} k^2\zeta^2K(\bk,\zeta)(\sum_{i=1,2}c_{i\bk}w^i +\sum_{j=3,4}d_{j\bk}w^j)e^{i\bk\cdot\bx}
\eeq

\section{Second Quantized Free Dirac Field}
\subsection{Heisenberg Picture}
Second quantization is fulfilled by anticommutators
$
\{c_{i\bk},c^\dag_{j\bk'}\}=\{d_{i\bk},d^\dag_{j\bk'}\}=\delta_{ij}\delta_{\bk,\bk'}
$
so that
\beq
\{\psi_A(x),\psi^\dag_B(y)\}_{|\Sigma}=\delta_{AB}\delta^3(Z).
\eeq
The second quantized Hamiltonian is accordingly
\begin{align}
H[\psi,\bpsi;\zeta]
&=\frac{1}{2}\frac{\ell^3}{\zeta^3}\sum_{\bk}A^2_{\bk}k^4\zeta^4 (c^\dag_{\bk},d^\dag_{\bk})\left(\begin{array}{cc}
\vep_1(\bk,\zeta)& 
\Delta(\bk,\zeta)\hat{\bk}\cdot\bsigma\\
 \Delta^*(\bk,\zeta)\hat{\bk}\cdot\bsigma& \vep_2(\bk,\zeta)
\end{array}
\right)
 \left(\begin{array}{c}c_{\bk}\\ d_{\bk}\end{array}\right)+h.c.
\end{align}
where ($z=k\zeta$)
\begin{align}
\vep_1(\bk,\zeta)&=e^{2\ell\pi m_\psi}[H^{(2)}_{\nu}(z)]^*\left[(\frac{3i}{2 \ell }+m_\psi ) H^{(2)}_{\nu}(z)+\frac{k\zeta}{\ell}i  H^{(2)}_{\nu-1}(z)\right]
                           -ie^{2\ell\pi m_\psi }[H^{(2)}_{\nu-1}(z)]^*\left[\frac{z}{\ell}H^{(2)}_\nu+(\frac{3i}{2\ell}-m_\psi)iH^{(2)}_{\nu-1}\right]\\
\vep_2(\bk,\zeta)&= [H^{(1)}_{\nu}(z)]^*\left[(\frac{3i}{2 \ell }+m_\psi )H^{(1)}_\nu(z)+i\frac{z}{\ell}H^{(1)}_{\nu-1}(z)\right]
        -i [H^{(1)}_{\nu-1}(z)]^*\left[\frac{z}{\ell}H^{(1)}_{\nu}+(\frac{3i }{2 \ell }-m_\psi )iH^{(1)}_{\nu-1}\right]\nonumber\\
\Delta(\bk,\zeta)&=-2iH^{(1)}_{\nu-1} H^{(1)}_{\nu}m_\psi+\frac{z}{\ell}( H^{(1)}_{\nu-1}H^{(1)}_{\nu-1} +H^{(1)}_{\nu}(z)H^{(1)}_\nu)\nonumber\\
\Delta^*(\bk,\zeta)
&=e^{2\ell\pi m_\psi}\left[2iH^{(2)}_{\nu} H^{(2)}_{\nu-1}m_\psi-\frac{z}{\ell}(H^{(2)}_{\nu}H^{(2)}_{\nu} +H^{(2)}_{\nu-1}(z)H^{(2)}_{\nu-1})\right]
\end{align}
Using the definition of Hankel functions, we have
\begin{align}
\vep_1(\bk,\zeta)+\vep^*_1(\bk,\zeta)   =2\vep     
 =-
(\vep_2(\bk,\zeta)+\vep^*_2(\bk,\zeta))
   \end{align}
with
\beq
\vep(\bk,\zeta)=e^{\ell\pi m_\psi }\Big[
   -i(H^{(1)}_{\nu-1} H^{(2)}_{\nu}+H^{(2)}_{\nu-1} H^{(1)}_{\nu})m_\psi+\frac{z}{\ell}( H^{(2)}_{\nu-1}H^{(1)}_{\nu-1} +H^{(1)}_{\nu}(z)H^{(2)}_\nu)
                              \Big] 
\eeq
and
\begin{align}
\left| \Delta \right|^2&=e^{2\ell\pi m_\psi}\Big[
4m_\psi^2 H^{(1)}_\nu H^{(1)}_{\nu-1}H^{(2)}_\nu H^{(2)}_{\nu-1}+2im_\psi\frac{z}{\ell} H^{(1)}_{\nu-1} H^{(1)}_{\nu} (H^{(2)}_{\nu}H^{(2)}_{\nu} +H^{(2)}_{\nu-1}(z)H^{(2)}_{\nu-1})\nonumber\\
&+2im_\psi\frac{z}{\ell} H^{(2)}_{\nu-1} H^{(2)}_{\nu} (H^{(1)}_{\nu}H^{(1)}_{\nu} +H^{(1)}_{\nu-1}(z)H^{(1)}_{\nu-1})\nonumber\\
&-(\frac{z}{\ell})^2 ( H^{(1)}_{\nu-1}H^{(1)}_{\nu-1} +H^{(1)}_{\nu}(z)H^{(1)}_\nu)
(H^{(2)}_{\nu}H^{(2)}_{\nu} +H^{(2)}_{\nu-1}(z)H^{(2)}_{\nu-1})\Big]
\end{align}
Since
\beq
\det \left(\begin{array}{cc}
\vep(\bk,\zeta)-\lambda& 
\Delta(\bk,\zeta)\hat{\bk}\cdot\bsigma\\
 \Delta^*(\bk,\zeta)\hat{\bk}\cdot\bsigma& -\vep(\bk,\zeta)-\lambda
\end{array}
\right)=\det \left(\vep(\bk,\zeta)-\lambda)(-\vep(\bk,\zeta)-\lambda)\MyI-\Delta(\bk,\zeta)
 \Delta^*(\bk,\zeta)\MyI\right)
 \eeq
 we find eigenvalues
\begin{align}
\lambda^2&=\vep^2(\bk,\zeta)+\left|\Delta\right|^2
=(m_\psi^2+\frac{z^2}{\ell^2})\frac{16}{\pi^2 z^2}
\end{align}
So the dispersion relations 
\beq
\omega_{\bk}(\zeta)=\pm\sqrt{m_\psi^2+\frac{z^2}{\ell^2}}
\eeq
which are the same as Klein-Gordon fields of {\it complementary series } obtained in \cite{Feng2020}.  Recalling that in Minkowski QFT, both Klein-Gordon field and Dirac field have the same dispersion relation. In de Sitter spacetime, after disentangling and unveiling all the intricacies of  first/second quantization and diagonalization, the resulting dispersion relations are still the same. This is not supposed to be considered coincidental to the authors. To diagonalize the Hamiltonian, we define $\alpha, \beta$ operators via
\beq
\left(\begin{array}{c}c_{\bk}\\d_{\bk}\end{array}\right)=M\left(\begin{array}{c}\alpha_{1\bk}\\\alpha_{2\bk}\\ \beta^\dag_{1-\bk}\\ \beta^\dag_{2-\bk}\end{array}\right)
\eeq
in which
\beq
M=\left(\begin{array}{cc} u_{\bk}(\zeta) &v_{\bk}(\zeta) \hat{\bk}\cdot\bsigma \\
-v^*_{\bk}(\zeta) \hat{\bk}\cdot\bsigma &u _{\bk}(\zeta) \end{array}\right)
\eeq
where $\hat{\bk}=\bk/k$ and
\begin{align}
u_{\bk}(\zeta)&=\frac{\omega_{\bk}(\zeta)+\vep(\bk,\zeta)}{\sqrt{[\omega_{\bk}(\zeta)+\vep(\bk,\zeta)]^2+|\Delta(\bk,\zeta)|^2}}\\
v_{\bk}(\zeta)&=\frac{-\Delta(\bk,\zeta)}{\sqrt{[\omega_{\bk}(\zeta)+\vep(\bk,\zeta)]^2+|\Delta(\bk,\zeta)|^2}}
\end{align}
Then
\beq
H(\zeta)=\sum_{\bk}\sum_{i=1}^2\omega_{\bk}(\zeta)\left[\alpha^\dag_{i\bk}(\zeta)\alpha_{i\bk}(\zeta)+\beta^\dag_{i\bk}(\zeta)\beta_{i\bk}(\zeta)\right]-2\sum_{\bk}\omega_{\bk}(\zeta)
\eeq
According to standard quantum theory of many-body systems\cite{Fetter1971}, $\alpha^\dag_{i\bk}(\zeta),\beta^\dag_{i\bk}(\zeta)$ generate observed quasi-particles/excitations. 
From the inverse
\begin{align}
\alpha_{\bk}(\zeta)=&u_{\bk}(\zeta) c_{\bk}-v_{\bk}(\zeta) \hat{\bk}\cdot\bsigma d_{\bk},\\
 \quad \beta^\dag_{-\bk}(\zeta)=&v_{\bk}^{*}(\zeta) \hat{\bk}\cdot\bsigma c_{\bk}+u_{\bk}(\zeta) d_{\bk},
\end{align}
we have the anticommutation relations
\begin{align}
\{\alpha_{i\bk}(\zeta_1),\alpha_{j\bp}^\dag(\zeta_2)\}=&(u_{\bk}(\zeta_1)u^*_{\bk}(\zeta_2)+v_{\bk}(\zeta_1)v^*_{\bk}(\zeta_2))\delta^3(\bk-\bp)\\
\{\alpha_{i\bk}(\zeta_1),\beta_{j-\bp}(\zeta_2)\}=&(u_{\bk}(\zeta_1)v_{-\bp}(\zeta_2)-v_{\bk}(\zeta_1)u_{-\bp}(\zeta_2))\hat{\bk}\cdot\bsigma_{ij}\delta^3(\bk-\bp)\\
\{\beta_{i-\bk}(\zeta_1),\beta_{j-\bp}^\dag(\zeta_2)\}=&(u_{\bk}(\zeta_1)u_{\bp}(\zeta_2)+v_{\bk}(\zeta_1)v^*_{\bp}(\zeta_2))\delta^3(\bk-\bp)
\end{align}
So $\alpha_{i\bk}(\zeta_1),\beta_{j-\bp}(\zeta_2)$ do not anticommute if $\zeta_1\not=\zeta_2$.
The momentum operata
\begin{align}
\bP&:=\sum_{\bk}\bk (c_{\bk}^\dag c_{\bk}+d^\dag_{\bk}d_{\bk})=\sum_{\bk}\sum_i\bk\left[\alpha^\dag_{i\bk}(\zeta)\alpha_{i\bk}(\zeta)+\beta^\dag_{i\bk}\beta_{i\bk}(\zeta)\right]
\end{align}
According to our previous definition, $p^a(\zeta)=(\omega_{\bk}(\zeta),C^{-1/2}(\zeta)\bk)$ is the measured 4-energy-momentum. It can be verified that this measurable energy-momentum satisfies the on-shell condition $p^a(\zeta)p_a(\zeta)=m^2_\psi$ and  the geodesic equation $p^\mu(\zeta)\nabla_\mu p^\nu(\zeta)=0$, which can be written as $p^\mu D_\mu p_a=0$, i.e. $p^\mu (\p_\mu p^a-\omega_{\mu }\,^a\,_bp^b)=0$. Recalling that  Erenfest's theorem proves the classical equation of motion of expected observables in quantum mechanics, this geodesic feature of measured energy-momentum can be considered playing a similar role in quantum field theory in curved spacetime.
\subsection{Schr\"{o}dinger Picture}
In Schr\"{o}dinger picture, we have to use $\xi$. The canonical fields are constant and agree with Heisenberg pictures fields at the chosen time $\zeta=\ell$. 
\beq
\xi^{\text{S}}(\bx)=\psi(\ell,\bx)=\sum_{\bk} A_{\bk} k^2\ell^2K(\bk,\ell)e^{i\bk\cdot\bx}\left(\begin{array}{c}c_{\bk}\\d_{\bk}\end{array}\right)
\eeq
The second-quantized Hamiltonian is then\begin{align}
&H^{\text{S}}[\xi^{\text{S}},\bar{\xi}^{\text{S}};\zeta]\nonumber\\
&=\frac{1}{2}\sum_{\bk}A^2_{\bk}k^4\ell^4 (c^\dag_{\bk},d^\dag_{\bk})\left(\begin{array}{cc}
\vep_1^{\text{S}}(\bk,\ell)& 
\Delta^{\text{S}}(\bk,\ell)\hat{\bk}\cdot\bsigma\\
 \Delta^{\text{S*}}(\bk,\ell)\hat{\bk}\cdot\bsigma& \vep_2^{\text{S}}(\bk,\ell)
\end{array}
\right)
 \left(\begin{array}{c}c_{\bk}\\ d_{\bk}\end{array}\right)+h.c.
\end{align}
$\vep^{\text{S}}_1(\bk,\ell)$ can be obtained from $\vep_1(\bk,\zeta)$ by replacement $\frac{3i}{2\ell}\rightarrow \frac{3i}{2\ell}+iC^{-1/2}(\zeta)\omega_0(\zeta)$ and $z=k\zeta\rightarrow k\ell$ in Hankel functions. But the factor $\frac{z}{\ell}$ in front of Hankel funcitons remain . In the calculation of $\vep_{1}+h.c.$, $\frac{3i}{2\ell}$ is canceled out and does not appear in the final expression. So we actually have
\beq
\vep^{\text{S}}(\bk,\zeta)=e^{\ell\pi m_\psi }\Big[
   -i(H^{(1)}_{\nu-1} (k\ell)H^{(2)}_{\nu}(k\ell)+H^{(2)}_{\nu-1} (k\ell)H^{(1)}_{\nu})(k\ell)m_\psi+\frac{k\zeta}{\ell}( H^{(2)}_{\nu-1}H^{(1)}_{\nu-1}(k\ell) +H^{(1)}_{\nu}(k\ell)H^{(2)}_\nu(k\ell))
                              \Big] 
\eeq
\begin{align}
\Delta(\bk,\zeta)
&=-2iH^{(1)}_{\nu-1} (k\ell)H^{(1)}_{\nu}(k\ell)m_\psi+\frac{k\zeta}{\ell}( H^{(1)}_{\nu-1}(k\ell)H^{(1)}_{\nu-1}(k\ell) +H^{(1)}_{\nu}(k\ell)H^{(1)}_\nu(k\ell))
\end{align}
 Hence
\begin{align}
H^{\text{S}}[\xi,\bar{\xi};\zeta]
&=\sum_{\bk}A^2_{\bk}k^4\ell^4 (c^\dag_{\bk},d^\dag_{\bk})\left(\begin{array}{cc}
\vep^{\text{S}}(\bk,\zeta)& 
\Delta^{\text{S}}(\bk,\zeta)\hat{\bk}\cdot\bsigma\\
 \Delta^{\text{S}*}(\bk,\zeta)\hat{\bk}\cdot\bsigma& -\vep^{\text{S}}(\bk,\zeta)
\end{array}
\right)\left(\begin{array}{c} c_{\bk}\\ d_{\bk}\end{array}\right)
\end{align}
So
\beq
\omega^{\text{S}}_{\bk}=\pm A_{\bk}^2 k^4\ell^4\frac{4}{\pi k\ell}\sqrt{m^2_\psi+\frac{z^2}{\ell^2}}=\omega_{\bk}(\zeta)
\eeq
Similar to Heisenberg picture, using
\beq
\left(\begin{array}{c}c_{\bk}\\d_{\bk}\end{array}\right)=M^{\text{S}}\left(\begin{array}{c}\alpha^{\text{S}}_{1\bk}\\\alpha^{\text{S}}_{2\bk}\\ \beta^{\text{S}\dag}_{1-\bk}\\ \beta^{\text{S}\dag}_{2-\bk}\end{array}\right) \label{CanoSch}
\eeq
where
\beq
M^{\text{S}}=\left(\begin{array}{cc} u_{\bk}^{\text{S}}(\zeta) &v_{\bk}^{\text{S}}(\zeta)\hat{\bk}\cdot\bsigma \\
-v_{\bk}^{\text{S}*}(\zeta)\hat{\bk}\cdot\bsigma &u_{\bk}^{\text{S}}(\zeta)\end{array}\right)
\eeq
and
\begin{align}
u^{\text{S}}_{\bk}(\zeta)&=\frac{\omega_{\bk}(\zeta)+\vep^{\text{S}}(\bk,\zeta)}{\sqrt{[\omega_{\bk}(\zeta)+\vep^{\text{S}}(\bk,\zeta)]^2+|\Delta^{\text{S}}(\bk,\zeta)|^2}}\\
v^{\text{S}}_{\bk}(\zeta)&=\frac{-\Delta^{\text{S}}(\bk,\zeta)}{\sqrt{[\omega_{\bk}(\zeta)+\vep^{\text{S}}(\bk,\zeta)]^2+|\Delta^{\text{S}}(\bk,\zeta)|^2}}
\end{align}
, then 
\beq
H^{\text{S}}(\zeta)=\sum_{\bk}\sum_{i=1}^2\omega_{\bk}(\zeta)\left[\alpha^{\text{S}^\dag}_{i\bk}(\zeta)\alpha^{\text{S}}_{i\bk}(\zeta)+\beta^{\text{S}^\dag}_{i\bk}(\zeta)\beta^{\text{S}}_{i\bk}(\zeta)\right]-2\sum_{\bk}\omega_{\bk}(\zeta)
\eeq
We definie $|\bO;\zeta\rj^{\text{S}}$ s.t.
\beq
 \alpha^{\text{S}}_{\bk}(\zeta)|\bO;\zeta\rj^{\text{S}}=\beta^{\text{S}}_{\bk}(\zeta)|\bO;\zeta\rj^{\text{S}}=0,\,\,\,\,\, \forall \,\,\,\,\ \bk,
\eeq
i.e.
\beq
|\bO;\zeta\rj^{\text{S}}=\prod_{\bk} |\bO_{\bk};\zeta\rj^{\text{S}} _{\bk}
\eeq
where
\beq
\alpha^{\text{S}}_{\bk} |\bO_{\bk};\zeta\rj_{\bk}^{\text{S}}=\beta^{\text{S}}_{\bk} |\bO_{\bk};\zeta\rj_{\bk}^{\text{S}}=0.
\eeq
Denoting $|0\rj$ as the state s.t. $c_{\bk}|0\rj=d^\dag_{\bk}|0\rj=0, \,\,\, \forall \bk$, since
\beq
\lj 0|H^{\text{S}}(\zeta)|0\rj=-2\sum_{\bk}\vep^{\text{S}}_{\bk}>\,^{\text{S}}\lj\bO;\zeta| H^{\text{S}}(\zeta)|\bO;\zeta\rj^{\text{S}}=-2\sum_{\bk}\omega_{\bk}(\zeta)
\eeq
,$|0\rj$ is therefore not the ground-state.  Here we have a system of fermionic time-dependent oscillators. As the case of Klein-Gordon field \cite{Feng2020}, the vacuum states are time-dependent.  The vacuum state at one time will evolve into a non-vacuum state at a later time. Thus particles can be generated.
 
 We now look for the relation between $|0\rj$ and $|\bO;\zeta\rj^{\text{S}}$.  The  following discussion holds for both Schr\"{o}dinger picture and Heisenberg pictures. Since (for convenience, denoting $f=d^\dag$).
 \beq
\left(\begin{array}{c}\alpha_{1\bk}\\\alpha_{2\bk}\\ \beta^{\dag}_{1-\bk}\\ \beta^{\dag}_{2-\bk}\end{array}\right) =M^{\dag} \left(\begin{array}{c}c_{\bk}\\f^\dag_{\bk}\end{array}\right)
\eeq
i.e., 
\beq
\alpha_{\bk}=u_{\bk}c_{\bk}-v_{\bk}\hat{\bk}\cdot\bsigma f^\dag_{\bk},\,\,\,\,\,\,
\beta^\dag_{-\bk}=v^*_{\bk}\hat{\bk}\cdot\bsigma c_{\bk}+u_{\bk}f^\dag_{\bk}
\eeq
 the canonical transformation eq.(\ref{CanoSch}) can be written as 
 \beq
\left(\begin{array}{c}b\\b^\dag\end{array}\right)
:=\left(\begin{array}{c}\alpha_{\bk}\\ \beta_{-\bk}\\ \alpha^\dag_{\bk}\\ \beta^{\dag}_{-\bk}\end{array}\right) =\left(\begin{array}{cccc}
u_{\bk}& 0 &0&-v_{\bk}\hat{\bk}\cdot\bsigma\\
0&u_{\bk}&v_{\bk}(\hat{\bk}\cdot\bsigma)^T&0\\
0& -v^*_{\bk}(\hat{\bk}\cdot \bsigma)^T&u_{\bk}&0\\
v^*_{\bk}\hat{\bk}\cdot\bsigma &0&0& u_{\bk}
\end{array}
\right)  \left(\begin{array}{c}a\\a^\dag\end{array}\right)
\eeq
in which
\beq
\left(\begin{array}{c}a\\a^\dag\end{array}\right)
=\left(\begin{array}{c}c_{\bk}\\f_{\bk}\\ c^\dag_{\bk}\\f^\dag_{\bk}\end{array}\right)
\eeq
Compared with \cite{Blaizot1985}, we have
\beq
T=\left(\begin{array} {cc}U&V\\ Y&X
\end{array}\right)
\eeq
with
\beq
U=\left(\begin{array} {cc}u_{\bk}&0\\ 0&u_{\bk}
\end{array}\right)=X,\,\,\,\,\,  V=\left(\begin{array} {cc}0&-v_{\bk}\hat{\bk}\cdot \bsigma\\ v_{\bk}(\hat{\bk}\cdot \bsigma)^T&0
\end{array}\right),\,\,\,\,\,\,
Y=\left(\begin{array} {cc}0&-v^*_{\bk}(\hat{\bk}\cdot \bsigma)^T\\ v^*_{\bk}\hat{\bk}\cdot \bsigma&0\end{array}\right)
\eeq
So
\beq
S=:\exp [-\frac{1}{2}a^\dag U^{-1}Va^\dag-\frac{1}{2} aYU^{-1}a+a^\dag(U^{-1}-1)a]:
\eeq
. Hence we have
\beq
|\bO;\zeta\rj=\hat{A}(\zeta)|0\rj=\otimes_{\bk}|\bO_{\bk};\zeta\rj_{\bk}
\eeq
where
\beq
\hat{A}(\zeta)=\mathpzc{A}(\zeta)\exp\left [\sum_{\bk}\frac{v_{\bk}}{u_{\bk}}(c^\dag_{1\bk}, c^\dag_{2\bk})(\hat{\bk}\cdot \bsigma)\left(\begin{array}{c}d_{1\bk}\\ d_{2\bk}\end{array}\right)\right]
\eeq
with
$
\mathpzc{A}(\zeta)=
\prod_{\bk}\mathpzc{A}_{\bk}(\zeta)
$
. Denoting
$
\gamma_{\bp}(\zeta)=\frac{v_{\bp}(\zeta)}{u_{\bp}(\zeta)}
$,
then
$
\hat{A}(\zeta)=\prod_{\bk}\hat{A}_{\bk}(\zeta)
$
with
$
\hat{A}_{\bk}(\zeta)=\mathpzc{A}_{\bk}(\zeta)\exp[\gamma_{\bk}(\zeta)(c^\dag_{1\bk}, c^\dag_{2\bk})(\hat{\bk}\cdot \bsigma)\left(\begin{array}{c}d_{1\bk}\\ d_{2\bk}\end{array}\right)]
$
. We need to find $\mathpzc{A}_{\bk}(\zeta)$ by
\begin{align}
|\mathpzc{A}_{\bk}(\zeta)|^{-2}&=\lj 0|   e^{\gamma^*_{\bk}(\zeta)(d^\dag_{1\bk}, d^\dag_{2\bk})(\hat{\bk}\cdot \bsigma)\left(\begin{array}{c}c_{1\bk}\\ c_{2\bk}\end{array}\right)}
e^{\gamma_{\bk}(\zeta)(c^\dag_{1\bk}, c^\dag_{2\bk})(\hat{\bk}\cdot \bsigma)\left(\begin{array}{c}d_{1\bk}\\ d_{2\bk}\end{array}\right)}|0\rj 
\end{align}
Hence we have $\mathpzc{A}_{\bk}(\zeta)=u^2_{\bk}(\zeta)$. 
We can calculate  $\lj\bO;\zeta_1| \bO;\zeta_2\rj$ as 
\begin{align}
\lj \bO;\zeta_1|\bO;\zeta_2\rj
=\prod_{\bk}[u_{\bk}(\zeta_1)u_{\bk}(\zeta_2)+v^*_{\bk}(\zeta_1)v_{\bk}(\zeta_2)]^2  \label{O|O}
\end{align}
By direct calculation, one can obtains the following expectations
\begin{align}
\lj\bO;\zeta''|c^\dag_{i\bk}c_{j\bp}|\bO;\zeta'\rj 
=&\delta_{\bk,\bp}\delta_{ij}v^*_{\bk}(\zeta'')v_{\bk}(\zeta')\big[u_{\bk}(\zeta'')u_{\bk}(\zeta')+v^*_{\bk}(\zeta'')v_{\bk}(\zeta')\big]
\label{cpc}\\
\lj\bO;\zeta''|d_{i\bk}d^\dag_{j\bp}|\bO;\zeta'\rj 
=&\delta_{\bk,\bp}\delta_{ij}v^*_{\bk}(\zeta'')v_{\bk}(\zeta')\big[u_{\bk}(\zeta'')u_{\bk}(\zeta')+v^*_{\bk}(\zeta'')v_{\bk}(\zeta')\big]
\\
\lj\bO;\zeta''|c^\dag_{i\bk}d^\dag_{j\bp}|\bO;\zeta'\rj 
=&0\\
 \lj\bO;\zeta''|c^\dag_{i\bk}d_{j\bp}|\bO;\zeta'\rj 
=&\delta_{\bk,\bp}\mathpzc{A}_{\bk}(\zeta'')\mathpzc{A}_{\bk}(\zeta')\cdot\gamma^*_{\bk}(\zeta'')(\hat{\bk}\cdot\bsigma)_{ji} 
[1+\gamma^*_{\bk}(\zeta'')\gamma_{\bk}(\zeta')]\label{pcd}
\end{align}
As is welknown in QFT, generation functions can be used to evaluate these expectations. Using the fermionic coherent states 
$ |z,w\rj=e^{c^\dag z+f^\dag w}|0\rj$, we have
\begin{align}
&\lj 0|e^{\gamma^*(\zeta'')f_i \sigma_{ij} c_j}e^{\lambda^*_ic_i}e^{ \mu^*_if_i}e^{f^\dag_i\mu_i }e^{c^\dag_i\lambda_i }e^{ \gamma(\zeta')c^\dag_i \sigma_{ij} f^\dag_j}|0\rj 
=\int dz^*dz dw^*dw  e^{W}
\end{align}
where
\beq
W=-z^*_iz_i-w_i^*w_i+\gamma^*(\zeta'')w_i \sigma_{ij} z_j+\lambda^*_iz_i+\mu^*_iw_i+w^*_i\mu_i+z^*_i\lambda_i 
+ \gamma(\zeta')z^*_i \sigma_{ij} w^*_j
\eeq
Letting $w^*=u$ and denoting
$
\chi=\left(\begin{array}{c}z_i\\u_i\end{array}\right), J=\left(\begin{array}{c}\lambda_i\\-\mu^*_i\end{array}\right)
$
, then $W$ can be written as 
\beq
W=\chi^\dag \left(\begin{array}{cc}
-\mathbb{I}_{2\times 2}& \gamma(\zeta')\sigma\\
\gamma^*(\zeta'')\sigma&\mathbb{I}_{2\times 2}
\end{array}\right)\chi+J^\dag \chi+\chi^\dag J
\eeq
Therefore
\begin{align}
&\lj 0|e^{\gamma^*(\zeta'')f_i \sigma_{ij} c_j}e^{\lambda^*_ic_i}e^{ \mu^*_if_i}e^{f^\dag_i\mu_i }e^{c^\dag_i\lambda_i }e^{ \gamma(\zeta')c^\dag_i \sigma_{ij} f^\dag_j}|0\rj =  \det \Theta e^{-J^\dag \Theta^{-1}J}=\det \Theta e^{-(1+\gamma^*(\zeta'')\gamma(\zeta'))^{-1}J^\dag \Theta J} \label{GZ}
\end{align}
where
\beq
\Theta=\left(\begin{array}{cc}
-\mathbb{I}_{2\times 2}& \gamma(\zeta')\sigma\\
\gamma^*(\zeta'')\sigma&\mathbb{I}_{2\times 2}
\end{array}\right)
\eeq 
Using eq.(\ref{cpc})-eq.(\ref{pcd}), one can calculate 
\begin{align}
\lj\bO;\zeta''|\alpha_{i\bk}(\zeta_1)\alpha_{j\bk}^\dag(\zeta_2)|\bO;\zeta'\rj 
&=\frac{\lj\bO;\zeta''|\bO;\zeta'\rj}{1+\gamma_{\bk}^*(\zeta'')\gamma_{\bk}(\zeta')}
\times [u_{\bk}(\zeta_1)+v_{\bk}(\zeta_1)\gamma^*_{\bk}(\zeta'')][u_{\bk}(\zeta_2)+v^*_{\bk}(\zeta_2)\gamma_{\bk}(\zeta')]\delta_{ij} \label{alphaalpha}
\\
\lj\bO;\zeta''|\beta_{i-\bk}(\zeta_1)\beta_{j-\bk}^\dag(\zeta_2)|\bO;\zeta'\rj 
&=\frac{\lj\bO;\zeta''|\bO;\zeta'\rj}{1+\gamma_{\bk}^*(\zeta'')\gamma_{\bk}(\zeta')}
\times [u_{\bk}(\zeta_1)+v_{\bk}(\zeta_1)\gamma^*_{\bk}(\zeta'')][u_{\bk}(\zeta_2)+v^*_{\bk}(\zeta_2)\gamma_{\bk}(\zeta')]\delta_{ij}
\\
\lj\bO;\zeta''|\alpha_{i\bk}(\zeta_1)\beta_{j-\bk}(\zeta_2)|\bO;\zeta'\rj 
 &=\frac{\lj\bO;\zeta''|\bO;\zeta'\rj}
{1+\gamma_{\bk}^*(\zeta'')\gamma_{\bk}(\zeta')} \hat{\bk}\cdot\bsigma_{ij}
 [u_{\bk}(\zeta_1)+v_{\bk}(\zeta_1)\gamma_{\bk}^*(\zeta'')][v_{\bk}(\zeta_2)-u_{\bk}(\zeta_2) \gamma_{\bk}(\zeta')]
\end{align}
 Expectations of more $\alpha, \beta$ operators can be evaluated by the following generation functionals.
\section{Generation Functional of Vacuum Expectations}
We define generation functional
\begin{align}
&\mathpzc{Z}[\theta,\theta^*, \rho, \rho^*  ;\zeta'',\zeta']
:=&\prod_{\bk}\, _{\bk}\lj \bO_{\bk};\zeta'' |
e^{\int d\eta\theta^\dag_{\bk}(\eta)\alpha_{\bk}(\eta)}e^{\int d\eta\alpha^\dag_{\bk}\theta_{\bk}(\eta)} e^{\int d\eta'\rho^\dag_{\bk}(\eta')\beta_{-\bk}(\eta')}e^{\int d\eta'\beta^\dag_{-\bk}(\eta')\rho_{\bk}(\eta')}|\bO_{\bk};\zeta'\rj_{\bk}
\end{align}
The $\theta_{\bk}(\eta), \theta^*_{\bk}(\eta), \rho_{\bk}(\eta), \rho^*_{\bk}(\eta)$ are external Grassmann $c-$fields and are independent of each other, instead of being mutually complex conjugate . For a particular $\bk$, we define
\begin{align}
\mathpzc{Z}_{\bk}[\theta,\theta^*, \rho, \rho^*  ;\zeta'',\zeta']
:=&_{\bk}\lj \bO_{\bk};\zeta'' |
e^{\int d\eta\theta^\dag_{\bk}(\eta)\alpha_{\bk}(\eta)}e^{\int d\eta\alpha^\dag_{\bk}\theta_{\bk}(\eta)} e^{\int d\eta'\rho^\dag_{\bk}(\eta')\beta_{-\bk}(\eta')}e^{\int d\eta'\beta^\dag_{-\bk}(\eta')\rho_{\bk}(\eta')}|\bO_{\bk};\zeta'\rj_{\bk}\nonumber\\
=&_{\bk}\lj \bO_{\bk};\zeta''| e^{\int d\eta\theta^*_{i\bk}(\eta)\left[u_{\bk}(\eta)c_{i\bk}-v_{\bk}(\eta)\hat{\bk}\cdot\bsigma_{im} d_{m\bk}\right]}e^{\int d\eta\left[u_{\bk}(\eta)c^\dag_{i\bk}-v^*_{\bk}(\eta)d^\dag_{m\bk}\hat{\bk}\cdot\bsigma_{mi} \right]\theta_{i\bk}(\eta)}\nonumber\\
&\times e^{\int d\eta'\rho^*_{j\bk}(\eta')\left[v_{\bk}(\eta')c^\dag_{n\bk}\hat{\bk}\cdot\bsigma_{nj} +u_{\bk}(\eta')d^\dag_{j\bk}\right]}e^{\int d\eta'\left[v^*_{\bk}(\eta')\hat{\bk}\cdot\bsigma_{jn} c_{n\bk}+u_{\bk}(\eta')d_{j\bk}\right]\rho_{j\bk}(\eta')}|\bO_{\bk};\zeta'\rj_{\bk}
\end{align}
Denoting
\begin{align}
\theta[u]=\int d\eta \theta_{\bk}(\eta)u_{\bk}(\eta),\,\,\,\,\,\,\,  \theta^*[v]=\int d\eta \theta^*_{\bk}(\eta)v_{\bk}(\eta),\,\,\,\,\,\,
\theta^*[u]=\int d\eta \theta^*_{\bk}(\eta)u_{\bk}(\eta),\,\,\,\,\,\,\,  \theta[v^*]=\int d\eta \theta_{\bk}(\eta)v^*_{\bk}(\eta)
\end{align}
then (omitting $\bk$ in $c, f, v,u,\theta, \rho,\gamma$ for notational convenience)
\begin{align}
&\mathpzc{Z}_{\bk}[\theta,\theta^*, \rho, \rho^*  ;\zeta'',\zeta']\nonumber\\
=&_{\bk}\lj \bO_{\bk};\zeta''| 
e^{\theta_i^*[u]c_i}
e^{-\theta_i^*[v]\hat{\bk}\cdot\bsigma_{im} f^\dag_m}
e^{-f_m(\hat{\bk}\cdot\bsigma_{mi} \theta_i[v^*]+\rho_m^*[u])}
e^{c^\dag_n(\theta_n[u]-\hat{\bk}\cdot\bsigma_{nj}\rho^*_j[v]) }
e^{-\rho_j[v^*]\hat{\bk}\cdot\bsigma_{jn} c_n}
e^{f^\dag_j\rho_j[u]}|\bO_{\bk};\zeta'\rj_{\bk}\nonumber\\
\end{align}
So
\begin{align}
\mathpzc{Z}_{\bk}[\theta,\theta^*, \rho, \rho^*  ;\zeta'',\zeta']
=&e^{-\theta_i^*[v]\hat{\bk}\cdot\bsigma_{im} \rho^*_m[u]-\theta_i^*[v]\theta_i[v^*]}
e^{-\rho_i[v^*]\hat{\bk}\cdot\bsigma_{in} \theta_n[u]+\rho_i[v^*]\rho^*_i[v]}\nonumber\\
&\times _{\bk}\lj \bO_{\bk};\zeta''| 
\exp\Big[\lambda^*_ic_i+\mu^*_if_i\Big]\exp\Big[c^\dag_i\lambda_i+f^\dag_i\mu_i\Big]|\bO_{\bk};\zeta'\rj_{\bk}
\end{align}
where
\begin{align}
\lambda^*_i&=\theta_i^*[u]-\rho_j[v^*]\hat{\bk}\cdot\bsigma_{ji})\\
\lambda_i&=\theta_i[u]-\hat{\bk}\cdot\bsigma_{ij}\rho^*_j[v]\\
\mu^*_m&=\hat{\bk}\cdot\bsigma_{mi} \theta_i[v^*]+\rho^*_m[u]\\
\mu_m&=\rho_m[u]+\theta_i^*[v]\hat{\bk}\cdot\bsigma_{im}
\end{align}
From eq.(\ref{GZ}),
we have
\begin{align}
&\mathpzc{Z}_{\bk}[\theta,\theta^*, \rho, \rho^*  ;\zeta'',\zeta']\nonumber\\
&=\mathpzc{A}(\zeta'')\mathpzc{A}(\zeta')e^{-\theta_i^*[v]\hat{\bk}\cdot\bsigma_{im} \rho^*_m[u]-\theta_i^*[v]\theta_i[v^*]}
e^{-\rho_i[v^*]\hat{\bk}\cdot\bsigma_{in} \theta_n[u]+\rho_i[v^*]\rho^*_i[v]}\det \Theta e^{-(1+\gamma^*(\zeta'')\gamma(\zeta'))^{-1}J^\dag \Theta J}
\end{align}
In particular, we have
\begin{align}
&\mathpzc{Z}_{\bk}[\theta,\theta^*, 0, 0 ;\zeta'',\zeta']=\mathpzc{A}(\zeta'')\mathpzc{A}(\zeta')e^{-\theta_i^*[v]\theta_i[v^*]}\det \Theta e^{-(1+\gamma^*(\zeta'')\gamma(\zeta'))^{-1}J^\dag \Theta J}
\end{align}
with
\begin{align}
J^\dag\Theta J&=-\theta_i^*[u] \theta_i[u]-\gamma^*(\zeta'')\theta_i^*[v]\theta_i[u]-\gamma(\zeta')\theta^*_i[u]\theta_i[v^*]+\theta^*_i[v]\theta_i[v^*]
\end{align}
and
$
\lambda^*_i=\theta_i^*[u],\,\,\,\,\,\lambda_i=\theta_i[u],\,\,\,\,\,\,\,\,
\mu^*_m=\hat{\bk}\cdot\bsigma_{mi} \theta_i[v^*],\,\,\,\,\,\,\,\,\mu_m=\theta_i^*[v]\hat{\bk}\cdot\bsigma_{im},\,\,\,\,\,\,\,\,\,
\mu_i\mu^*_i=\theta_i^*[v] \theta_i[v^*]
$
. The  previous  amplitudes eq.(\ref{alphaalpha}). can be therefore evaluated.
\section{Transition of States of Free Dirac Field}
 As in conventional quantum field theories\cite{Itzykson1980}, suppose that at time $\zeta_1$, the system is in a vacuum state, the state at $\zeta_2$ has some computable probability to contain multiple particles. In Schr\"{o}dinger picture, the state
\begin{align}
|\Psi_0(\zeta_2;\zeta_1)\rj^{\text{S}}=\hat{T}e^{-i\int^{\zeta_2}_{\zeta_1}H^{\text{S}}[\xi(\bx),\bar{\xi}(\bx);\eta]e^0_\zeta(\eta)d\eta}|\bO;\zeta_1\rj ^{\text{S}}
\end{align}
is a formal solution to
\beq
i\hat{\nabla}_0|\Psi_0(\zeta;\zeta_1)\rj^{\text{S}}=H^{\text{S}}[\xi(\bx),\bar{\xi}(\bx);\zeta] |\Psi_0(\zeta;\zeta_1)\rj ^{\text{S}}
\eeq
with initial condition $|\Psi_0(\zeta_1;\zeta_1)\rj ^{\text{S}}=|\bO;\zeta_1\rj ^{\text{S}}$.
The transition amplitude from state $|\{n^\alpha_{\bk},n^\beta_{-\bk}\};\zeta_1\rj^{\text{S}}$ to state  $|\{m^\alpha_{\bk},m^\beta_{-\bk}\};\zeta_2\rj ^{\text{S}}$ is given by
\begin{align}
&\mathscr{T}(|\{n^\alpha_{\bk},n^\beta_{-\bk}\};\zeta_1\rj ^{\text{S}}\rightarrow |\{m^\alpha_{\bk},m^\beta_{-\bk}\};\zeta_2\rj ^{\text{S}})\nonumber\\
=&\,^{\text{S}}\lj\{m^\alpha_{\bk},m^\beta_{-\bk}\};\zeta_2| \hat{T}e^{-i\int^{\zeta_2}_{\zeta_1}H^{\text{S}}(\eta)e^0_\zeta(\eta)d\eta}|\{n^\alpha_{\bk},n^\beta_{-\bk}\};\zeta_1\rj ^{\text{S}}
\end{align}
For a particular $\bk$, 
\begin{align}
&\mathscr{T}(|n^\alpha_{\bk},n^\beta_{-\bk};\zeta_1\rj_{\bk}^{\text{S}}\rightarrow |m^\alpha_{\bk},m^\beta_{-\bk};\zeta_2\rj_{\bk}^{\text{S}})\nonumber\\
&=\,^{\text{S}}_{\bk}\lj m^\alpha_{\bk},m^\beta_{-\bk};\zeta_2|\hat{T}e^{-i\int^{\zeta_2}_{\zeta_1}\omega_{\bk}(\eta)\left[\alpha^{\text{S}\dag}_{\bk}(\eta)\alpha^{\text{S}}_{\bk}(\eta)+
\beta^{\text{S}\dag}_{-\bk}(\eta)\beta^{\text{S}}_{-\bk}(\eta)-2\right]e^0_\zeta(\eta) d\eta}\nonumber\\
&\times |n^\alpha_{\bk},n^\beta_{-\bk};\zeta_1\rj_{\bk} ^{\text{S}}
\end{align}
Slicing the time as 
$\eta_0=\zeta_1, \eta_{j}=\eta_{j-1}-\varDelta,\varDelta=\frac{\zeta_1-\zeta_2}{N}, \eta_N=\zeta_2$. 
\begin{align}
&\mathscr{T}(|n^\alpha_{\bk},n^\beta_{-\bk};\zeta_1\rj_{\bk}^{\text{S}}\rightarrow |m^\alpha_{\bk},m^\beta_{-\bk};\zeta_2\rj_{\bk}^{\text{S}})\nonumber\\
=&\lim_{N\rightarrow\infty}\, ^{\text{S}}_{\bk}\lj m^\alpha_{\bk},m^\beta_{-\bk};\zeta_2|\nonumber\\
&\times
e^{-i\omega_{\bk}(\eta_{N-1})\left[\alpha^{\text{S}\dag}_{\bk}(\eta_{N-1})\alpha^{\text{S}}_{\bk}(\eta_{N-1})+\beta^{\text{S}\dag}_{-\bk}(\eta_{N-1})\beta^{\text{S}}_{-\bk}(\eta_{N-1})-2\right]e^0_\zeta(\eta_{N-1}) \varDelta}\nonumber\\
&\times
e^{-i\omega_{\bk}(\eta_{N-2})\left[\alpha^{\text{S}\dag}_{\bk}(\eta_{N-2})\alpha^{\text{S}}_{\bk}(\eta_{N-2})+\beta^{\text{S}\dag}_{-\bk}(\eta_{N-2})\beta^{\text{S}}_{-\bk}(\eta_{N-2})-2\right]e^0_\zeta(\eta_{N-2}) \varDelta}
\nonumber\\
&\times \cdots
e^{i\omega_{\bk}(\eta_{1})\left[\alpha^{\text{S}\dag}_{\bk}(\eta_{1})\alpha^{\text{S}}_{\bk}(\eta_{1})+\beta^{\text{S}\dag}_{-\bk}(\eta_{1})\beta^{\text{S}}_{-\bk}(\eta_{1})-2\right]e^0_\zeta(\eta_{1})\varDelta }\nonumber\\
&\times 
e^{-i\omega_{\bk}(\eta_{0})\left[\alpha^{\text{S}\dag}_{\bk}(\eta_{0})\alpha^{\text{S}}_{\bk}(\eta_{0})+\beta^{\text{S}\dag}_{-\bk}(\eta_{0})\beta^{\text{S}}_{-\bk}(\eta_{0})-2\right]e^0_\zeta(\eta_{0})\varDelta }
|n^\alpha_{\bk},n^\beta_{-\bk};\zeta_1\rj_{\bk}^{\text{S}}\nonumber\\
=&\lim_{N\rightarrow\infty}\, ^{\text{S}}_{\bk}\lj m^\alpha_{\bk},m^\beta_{-\bk};\zeta_2|e^{2i\int^{\zeta_2}_{\zeta_1}\omega_{\bk}(\eta)e^0_{\zeta}(\eta)d\eta}\nonumber\\
&\times
e^{-i\omega_{\bk}(\eta_{N-1})\left[\alpha^{\text{S}\dag}_{\bk}(\eta_{N-1})\alpha^{\text{S}}_{\bk}(\eta_{N-1})+\beta^{\text{S}\dag}_{-\bk}(\eta_{N-1})\beta^{\text{S}}_{-\bk}(\eta_{N-1})\right]e^0_\zeta(\eta_{N-1}) \varDelta}\nonumber\\
&\times
e^{-i\omega_{\bk}(\eta_{N-2})\left[\alpha^{\text{S}\dag}_{\bk}(\eta_{N-2})\alpha^{\text{S}}_{\bk}(\eta_{N-2})+\beta^{\text{S}\dag}_{-\bk}(\eta_{N-2})\beta^{\text{S}}_{-\bk}(\eta_{N-2})\right]e^0_\zeta(\eta_{N-2}) \varDelta}
\nonumber\\
&\times \cdots
e^{i\omega_{\bk}(\eta_{1})\left[\alpha^{\text{S}\dag}_{\bk}(\eta_{1})\alpha^{\text{S}}_{\bk}(\eta_{1})+\beta^{\text{S}\dag}_{-\bk}(\eta_{1})\beta^{\text{S}}_{-\bk}(\eta_{1})\right]e^0_\zeta(\eta_{1})\varDelta }\nonumber\\
&\times 
e^{-i\omega_{\bk}(\eta_{0})\left[\alpha^{\text{S}\dag}_{\bk}(\eta_{0})\alpha^{\text{S}}_{\bk}(\eta_{0})+\beta^{\text{S}\dag}_{-\bk}(\eta_{0})\beta^{\text{S}}_{-\bk}(\eta_{0})\right]e^0_\zeta(\eta_{0})\varDelta }
|n^\alpha_{\bk},n^\beta_{-\bk};\zeta_1\rj_{\bk}^{\text{S}}
\end{align}
Using coherent states $j=0,1,\cdots, N$
\beq
|z_{j\bk}, z_{j-\bk}\rj=e^{z_{j\bk}\alpha^{\text{S}\dag}_{\bk}(\eta_j)+z_{j-\bk}\beta^{\text{S}\dag}_{-\bk}(\eta_j)}|\bO_{\bk};\eta_j\rj_{\bk}
\eeq
\begin{align}
&\mathscr{T}(|n^\alpha_{\bk},n^\beta_{-\bk};\zeta_1\rj_{\bk}^{\text{S}}\rightarrow |m^\alpha_{\bk},m^\beta_{-\bk};\zeta_2\rj_{\bk}^{\text{S}})\nonumber\\
&=\lim_{N\rightarrow\infty}e^{2i\int^{\zeta_2}_{\zeta_1}\omega_{\bk}(\eta)e^0_{\zeta}(\eta)d\eta}
\int \left[\prod_{j=0}^Ndz_{j\bk}dz^*_{j\bk} \right]\nonumber\\
&\times  z^{m_{\bk}}_{N\bk} z^{*n_{\bk}}_{0\bk}
\exp\left[-z^*_{0\bk}z_{0\bk}\right]\nonumber\\
&\times
\prod_{j=1}^{N}
\exp\left[-z^*_{j\bk}z_{j\bk}+z^*_{j\bk}z_{j-1,\bk}-i\omega_{\bk}(\eta_{j-1})z^*_{j\bk}z_{j-1\bk}e^0_\zeta(\eta_{j-1}) \varDelta\right]\nonumber\\
&\times
\int \left[\prod_{j=0}^N dz_{j-\bk}dz^*_{j-\bk}\right]\,
 z^{m_{-\bk}}_{N-\bk}z^{*n_{-\bk}}_{0-\bk}
\exp\left[-z^*_{0-\bk}z_{0-\bk}\right]\nonumber\\
&\times
\prod_{j=1}^{N}
\exp\Big[-z^*_{j-\bk}z_{j-\bk}+z^*_{j-\bk}z_{j-1,-\bk}\nonumber\\
&-i\omega_{\bk}(\eta_{j-1})z^*_{j-\bk}z_{j-1,-\bk}e^0_\zeta(\eta_{j-1}) \varDelta\Big]
\end{align}
which can be written as a path-integral by shorthand. It is easy to see that starting from a vacuum state at $\zeta_1$, the state will evolve into a mixed states at later time $\zeta_2$, which is not unusual for systems in external fields in  Minkowski quantum field theories\cite{Itzykson1980} and time-dependent harmonic oscillators\cite{Chernikov1967}-\cite{Struckmeier2001}.
\end{widetext}

\section{Perturbation}
In parallel with the discussion of $\lambda^4$-theory in\cite{Feng2020}, we present a formal discussion of an interacting Dirac field. 
\beq
\mathscr{L}=\mathscr{L}_0+\mathscr{L}_{\text{int}}(\bpsi,\psi;e^\mu_a)
\eeq
where $\mathscr{L}_0$ is the Lagrangian of free Dirac field eq.(\ref{symL}) and $\mathscr{L}_0$ is the interacting Lagrangian such as current-current interaction. The full Hamiltonian is
\begin{align}
H[\psi,\bpsi;\zeta]=&H_0[\psi,\bpsi;\zeta]+H_{\text{I}}[\psi,\bpsi;\zeta]
\end{align}
of which $\mathscr{H}_0$ is the Hamiltonian eq.(\ref{phyH}) of free Dirac field and 
$
\mathscr{H}_{\text{I}}=-\mathscr{L}_{\text{int}}(\bpsi,\psi;e^\mu_a).
$
 In terms of fields $\xi,\bar{\xi}$, the Schr\"{o}dinger state follows
 \beq
 i\hat{\nabla}_0|\Psi(\zeta)\rj^{\text{S}}=H[\xi^{\text{S}}(\bx),\xi^{\text{S}}(\bx);\zeta] |\Psi(\zeta)\rj^{\text{S}}
  \eeq
where the free part of $H[\xi^{\text{S}}(\bx),\xi^{\text{S}}(\bx);\zeta]$ is defined as in eq.(\ref{xiH}) and the interacting part is supposed to be appropriate substitution of $\xi,\bar{\xi}$ into $\mathscr{L}_{\text{I}}$. Defining Dirac picture state  
 \beq
|\Psi(\zeta)\rj^{\text{D}}:=\hat{T}^{-1}e^{i\int_\ell^{\zeta}H_0^{\text{S}}(\eta)e^0_\zeta(\eta)d\eta} |\Psi(\zeta)\rj^{\text{S}}
  \eeq  
  hence the two pictures coincide at time $\zeta=\ell,
|\Psi(\ell)\rj^{\text{S}}=|\Psi(\ell)\rj^{\text{D}}
$
  we thus have equation of motion 
 \begin{align}
 i\hat{\nabla}_0| \Psi(\zeta)\rj^{\text{D}}
=&H^{\text{D}}_{\text{I}}(\zeta)  |\Psi(\zeta)\rj^{\text{D}}
  \end{align} 
  where
\begin{align}
 H^{\text{D}}_{\text{I}}(\zeta)
=&U^{\text{S}-1}_0(\zeta,\ell) H_{\text{I}}^{\text{S}}(\zeta)U^{\text{S}}_0(\zeta,\ell)
\end{align}
with
\beq
U^{\text{S}}_0(\zeta'',\zeta')=\hat{T} e^{-i\int^{\zeta''}_{\zeta'} H^{\text{S}}_0(\eta)e^0_\zeta(\eta)d\eta}
\eeq
The Dirac picture field and operators are defined
\begin{align}
\xi^{\text{D}}(\zeta,\bx):=&U^{\text{S}-1}_0(\zeta,\ell)\xi^{\text{S}}(\bx)U^{\text{S}}_0(\zeta,\ell)\\
H_0^{\text{D}}(\zeta):=&U^{\text{S}-1}_0(\zeta,\ell) H_0^{\text{S}}(\zeta)U^{\text{S}}_0(\zeta,\ell) \label{HD}
\end{align}
, from which it follows that
\begin{align}
ie_0^{\zeta}(\zeta)\frac{\p}{\p\zeta} \xi^{\text{D}}(\zeta,\bx)=&[\xi^{\text{D}}(\zeta,\bx),H^{\text{D}}_0]\\
ie_0^{\zeta}(\zeta)\frac{\p}{\p\zeta} \bar{\xi}^{\text{D}}(\zeta,\bx)=&[\bar{\xi}^{\text{D}}(\zeta,\bx),H^{\text{D}}_0]
\end{align}
Hence $\xi^{\text{D}}(\zeta,\bx),\bar{\xi}^{\text{D}}(\zeta,\bx)$ follow equation of motion of a non-interacting field, of which the time-dependence was discussed previously.
Defining
\begin{align}
U^{\text{D}}_0(\zeta'',\zeta')=&\hat{T} e^{-i\int^{\zeta''}_{\zeta'} H^{\text{D}}_0(\eta)e^0_\zeta(\eta)d\eta}
\end{align}
we have
\begin{align}
U^{\text{D}\dag}_0(\zeta'',\zeta')=&\hat{T}^{-1} e^{i\int^{\zeta''}_{\zeta'} H^{\text{D}}_0(\eta)e^0_\zeta(\eta)d\eta}=U^{\text{D}-1}_0(\zeta'',\zeta')
\end{align}
Further, by definition eq.(\ref{HD}) of $H_0^{\text{D}}(\zeta)$, we have
\beq
H_0^{\text{D}}(\zeta)e^0_\zeta(\zeta)=U^{\text{S}-1}_0(\zeta,\ell) i\p_\zeta U^{\text{S}}_0(\zeta,\ell)
\eeq
Therefore
\beq
i\p_\zeta U^{\text{S}-1}_0(\zeta,\ell)=-H_0^{\text{D}}(\zeta)e^0_\zeta(\zeta) U^{\text{S}-1}_0(\zeta,\ell)
\eeq
Hence we have the following relations
\begin{align}
U^{\text{S}-1}_0(\zeta,\ell)= &\hat{T}e^{i\int^{\zeta}_{\ell} H^{\text{D}}_0(\eta)e^0_\zeta(\eta)d\eta}\\
U^{\text{S}}_0(\zeta,\ell)= &\hat{T}^{-1}e^{-i\int^{\zeta}_{\ell} H^{\text{D}}_0(\eta)e^0_\zeta(\eta)d\eta}\\
\hat{T}e^{-i\int_\ell^{\zeta}H_0^{\text{S}}(\eta)e^0_\zeta(\eta)d\eta}=&\hat{T}^{-1}e^{-i\int_\ell^{\zeta}H_0^{\text{D}}(\eta)e^0_\zeta(\eta)d\eta}
\end{align}
we have inverse transformation
\beq
H_0^{\text{S}}(\zeta)=\hat{T}^{-1}e^{-i\int_\ell^{\zeta}H_0^{\text{D}}(\eta)e^0_\zeta(\eta)d\eta} H_0^{\text{D}}(\zeta)\hat{T}e^{i\int_\ell^{\zeta}H_0^{\text{D}}(\eta)e^0_\zeta(\eta)d\eta}
\eeq
and
\beq
| \Psi(\zeta)\rj^{\text{S}}
=\hat{T}^{-1}e^{-i\int_\ell^{\zeta}H_0^{\text{D}}(\eta)e^0_\zeta(\eta)d\eta}| \Psi(\zeta)\rj^{\text{D}}
  \eeq  

Suppose at the initial time $\zeta=\ell$, the system is in the eigen-state $|A;\ell\rj_0^{\text{S}}$ of $H_0^{\text{S}}(\ell)$, then interaction is turned on adiabatically.
At time $\zeta=0$, the interaction is turned off and the state evolves  into a state which can be expanded in terms of eigen-states $\{|B;0\rj_0^{\text{S}}\} $ of $H_0^{\text{S}}(0)$. The probability of the transition is the square of the amplitude $^{\text{S}}_0\lj B;0| U^{\text{S}}(0,\ell)|A,\ell\rj_0^{\text{S}}$
 \cite[p.323]{Merzbacher1998}, where the Schr\"{o}dinger picture evolution operator is
\beq
U^{\text{S}}(\zeta_2,\zeta_1)=\hat{T}e^{-i\int_{\zeta_1}^{\zeta_2}H^{\text{S}}(\eta)e^0_{\zeta}(\eta)d\eta}
\eeq
For a free field, we have
\beq
^{\text{S}}_0\lj B;0| U^{\text{S}}(0;\ell)|A,\ell\rj_0^{\text{S}}=\,^{\text{S}}_0\lj B;0| \hat{T}e^{-i\int_\ell^{0}H_0^{\text{S}}(\eta)e^0_{0}(\eta)d\eta}|A,\ell\rj_0^{\text{S}}
\eeq
, which was discussed previously. In the interacting case, in terms of Dirac picture
\beq
| \Psi(\zeta)\rj^{\text{D}}
=\hat{T}e^{-i\int_\ell^{\zeta}H^{\text{D}}_{\text{I}}(\eta)e^0_{\zeta}(\eta)d\eta}  |\Psi(\ell)\rj^{\text{D}}
  \eeq  
\begin{align}
&| \Psi(\zeta)\rj^{\text{S}}\nonumber\\
=&\hat{T}^{-1}e^{-i\int_\ell^{\zeta}H_0^{\text{D}}(\eta)e^0_\zeta(\eta)d\eta}\hat{T}e^{-i\int_\ell^{\zeta}H^{\text{D}}_{\text{I}}(\eta)e^0_{\zeta}(\eta)d\eta}  |\Psi(\ell)\rj^{\text{S}}
\end{align}
Hence we have expression of Schr\"{o}dinger picture evolution operator using only Dirac picture operators
\begin{align}
U^{\text{S}}(\zeta,\ell)
=&\hat{T}^{-1}e^{-i\int_\ell^{\zeta}H_0^{\text{D}}(\eta)e^0_\zeta(\eta)d\eta}\hat{T}e^{-i\int_\ell^{\zeta}H^{\text{D}}_{\text{I}}(\eta)e^0_{\zeta}(\eta)d\eta}  
\end{align}
and
$
| \Psi(\zeta)\rj^{\text{S}}
=U^{\text{S}}(\zeta,\ell)|\Psi(\ell)\rj^{\text{S}}.
$
We have the transition amplitude  \cite[p.484]{Merzbacher1998}  
\begin{align}
&^{\text{S}}_0\lj B;0| U^{\text{S}}(0,\ell)|A;\ell\rj_0^{\text{S}}\nonumber\\
=&\,^{\text{S}}_0\lj B;0|\hat{T}e^{-i\int_\ell^{0}H_0^{\text{S}}(\eta)e^0_\zeta(\eta)d\eta}\hat{T}e^{-i\int_\ell^{0}H^{\text{D}}_{\text{I}}(\eta)e^0_{0}(\eta)d\eta} |A;\ell\rj_0^{\text{S}}
\end{align}
This is the basis for perturbational calculations since the second factor can be expanded in terms of powers of interacting parameter in $\mathscr{L}_{\text{I}}$.  In this relation, dependence of fields in  $H^{\text{D}}_{\text{I}}$ on time is the same as in the Heisenberg fields in the non-interacting  case while $H^{\text{S}}_0$ is the same as in Schr\"{o}dinger fields in the non interacting case. $H^{\text{D}}_{\text{I}}$ is supposed to be expressed in terms of $\alpha_{\bk}(\zeta)\alpha^\dag_{\bk}(\zeta),  \beta_{\bk}(\zeta), \beta^\dag_{\bk}(\zeta)$ while $H^{\text{S}}$  in terms of $\alpha^{\text{S}}_{\bk}(\zeta),\alpha^{\text{S}\dag}_{\bk}(\zeta),  \beta^{\text{S}}_{\bk}(\zeta),\beta^{\text{S}\dag}_{\bk}(\zeta)$.  These two set of quasi-particles operators are related to $c_{\bk},c^\dag_{\bk},d_{\bk}, d^\dag_{\bk}$ in two different ways.

\section{Local Lorentz Covariance}
For free Klein-Gordon field, we can define, using the stress-energy tensor
\beq
T_{\mu\nu}=\nabla_\mu\phi\nabla_\nu\phi-g_{\mu\nu}\mathscr{L}_\phi
\eeq
that
\begin{align}
H_a&=\int_{\Sigma} d\sigma^\mu e^\nu_aT_{\mu\nu}=\int_{\Sigma}d\sigma^\zeta e^b_\zeta( \hat{\nabla}_b\phi\hat{\nabla}_a\phi-\eta_{ab}\mathscr{L}_\phi)
\end{align}
then ($d\sigma^\zeta=g^{0\mu}d\sigma_\mu=C^{-1} C^2 d^3\bx, e^b_\zeta=\delta^b_0 C^{1/2}, \dot{g}^*_{\bk}g_{\bk}-\dot{g}_{\bk}g^*_{\bk}=i(2\pi)^{-3}\ell^{-2}\zeta^2 .$)\cite{Feng2020} 
\begin{align}
H_0&=H\\
H_{a'}&=\int_{\Sigma}d\sigma^\zeta e^b_\zeta \hat{\nabla}_b\phi\hat{\nabla}_{a'}\phi&=C^{-1/2}(\zeta)\delta^i_{a'}\sum_{\bk}k_{i}a^\dag_{\bk}a_{\bk}
\end{align}
Hence we have local Lorentz covariant Heisenberg equation
\beq
i\hat{\nabla}_a\phi=[\phi,H_a]
\eeq
and $H_a$  is precisely the {\it measured} 4-energy-momentum in frame $e_{\mu a}$. If the quantization is implemented using a different Lorentz gauge, $e'_{\mu a}=\Lambda_a\,^b(x) e_{\mu b}$, the resulting $H_a'$ will be different and relation $H'_a=\Lambda_a\,^b H_b$ holds only when $\Lambda$ is a global transformation. So in general, $H_a$ will not be the measured energy-momentum by a local observer at $x$. Instead, the measured energy-momentum of a quanta  should be boosted by $\Lambda(x)$, i.e., $\Lambda_a\,^b(x)H_b$. Therefore, though the quantization can be implemented in any Lorentz gauge, the measured local physical quantities should be related to a particular choice of vierbein. This is expected\cite{Feng1997}-\cite{Feng2001}.

For free Dirac field, we use  
\beq
\mathscr{T}_{\mu\nu}=\frac{i}{2}(\bpsi \gamma_ae^a_\mu\DD_\nu\psi-\bpsi \overleftarrow{\DD}_\nu\gamma_ae^a_\mu\psi)-g_{\mu\nu}\mathscr{L}_\psi
\eeq    
 and define
 \beq
 H_a=\int_{\Sigma}d\sigma^\mu e^\nu_a \mathscr{T}_{\mu\nu}
 \eeq
Then we have
\beq
H_0=H_{\text{phy}}
\eeq
and
\begin{align}
    H_{a'}
&=\frac{i}{2}\int_{\Sigma} d\sigma^\zeta( \bpsi \gamma_be^b_\zeta e^i_{a'}\DD_i\psi-\bpsi \overleftarrow{\DD}_i\gamma_be^b_\zeta e^i_{a'}\psi)\nonumber\\
&=C^{-1/2}(\zeta)\sum_{\bk} \delta^i_{a'}k_i (\alpha^\dag_{\bk}(\zeta)\alpha_{\bk}(\zeta)+\beta_{-\bk}(\zeta)\beta^\dag_{-\bk}(\zeta))
\end{align}    
Using
\beq
\bpsi\overleftarrow{\DD}_\nu e^b_\mu \gamma_b\psi=\p_\nu(\bpsi e^b_\mu\gamma_b\psi)-\bpsi e^b_\mu\gamma_b\DD_\nu\psi
\eeq
we have
 \beq
   i\hat{\DD}_a\psi=[\psi, H_a]
   \eeq
 , which shows both general covariance and local Lorentz covariance.

\section{Discussions}
General relativity and QFT are two pillars of modern theoretical physics.  As a preamble of a complete unified quantum theory of  gravity and matter systems, quantum field theories in classical curved spacetimes have long been called for. 
Following our previous work, we here proposed a generally covariant framework for quantizing Dirac field in de Sitter spacetime. The framework is formulated in conformal coordinate which is specifically chosen. It can be transformed into other coordinate systems $x'$. The fundamental solutions will still be labelled by quantum numbers $\underline{k}$ but the functions will take a more complex appearance depending on the coordinates $x'$. The surfaces $\Sigma$ will be defined by functions $\zeta=\zeta(x')=\text{Const.}$  In the new coordinate system $x'$, the {\it time-dependence} becomes actually $\Sigma$-{\it dependence}. 

As our previous work on Klein-Gordon field, our framework of quantizing Dirac field is covariant under both general coordinates transformations and Local Lorentz transformations. This framework provides many quantum concepts in parallel with the standard quantum field theories in Minkowski spacetime. 
Our approach is fundamentally different from other discussions in the literatures. Two features show consistency of our approach with standard QFT in Minkowski spacetime and general relativity. The energy of free particles are the same as Klein-Gordon fields. The measurable energy-momentum 4-vector satisfies geodesic equation. 

 Our framework also enjoys the three traditional pictures: Heisenberg, Schr\"{o}dinger and Dirac in an extended fashion. The Hamiltonians in Heisenberg and Schr\"{o}dinger pictures are not identical anymore and  so are not the non-interacting Hamiltonians of Dirac and Schr\"{o}dinger picture equal.  Yet, we can nevertheless devise a way to calculate perturbatively the impact of interaction provided the coupling is weak.

\end{document}